\documentclass[preprint]{aastex63}

\usepackage{lineno}
\revised{February 2 2022}
\submitjournal{AJ}

\shorttitle{Handling Background in IXPE polarimetric data}
\shortauthors{Di Marco et al.}

\begin{document}

\title{Handling the background in IXPE polarimetric data}

\correspondingauthor{Alessandro Di Marco}
\email{alessandro.dimarco@inaf.it}

\author[0000-0003-0331-3259]{Alessandro Di Marco}
\affiliation{INAF -- IAPS, via Fosso del Cavaliere, 100, Rome, Italy I-00133}

\author[0000-0002-7781-4104]{Paolo Soffitta}
\affiliation{INAF -- IAPS, via Fosso del Cavaliere, 100, Rome, Italy I-00133}

\author[0000-0003-4925-8523]{Enrico Costa}
\affiliation{INAF -- IAPS, via Fosso del Cavaliere, 100, Rome, Italy I-00133}

\author[0000-0003-1074-8605]{Riccardo Ferrazzoli}
\affiliation{INAF -- IAPS, via Fosso del Cavaliere, 100, Rome, Italy I-00133}

\author[0000-0001-8916-4156]{Fabio La Monaca}
\affiliation{INAF -- IAPS, via Fosso del Cavaliere, 100, Rome, Italy I-00133}

\author[0000-0002-9774-0560]{John Rankin}
\affiliation{INAF -- IAPS, via Fosso del Cavaliere, 100, Rome, Italy I-00133}

\author[0000-0003-0411-4243]{Ajay Ratheesh}
\affiliation{INAF -- IAPS, via Fosso del Cavaliere, 100, Rome, Italy I-00133}

\author[0000-0002-0105-5826]{Fei Xie}
\affiliation{Guangxi Key Laboratory for Relativistic Astrophysics, School of Physical Science and Technology,Guangxi University, Nanning, China}
\affiliation{INAF -- IAPS, via Fosso del Cavaliere, 100, Rome, Italy I-00133}

\author[0000-0002-9785-7726]{Luca Baldini}
\affiliation{Universit\`a di Pisa, Dipartimento di Fisica Enrico Fermi, Largo B. Pontecorvo 3, Pisa, Italy, I-56127}
\affiliation{INFN-Pisa, Largo B. Pontecorvo 3, Pisa, Italy, I-56127}

\author[0000-0002-3013-6334]{Ettore Del Monte}
\affiliation{INAF -- IAPS, via Fosso del Cavaliere, 100, Rome, Italy I-00133}

\author[0000-0003-4420-2838]{Steven R. Ehlert}
\affiliation{NASA Marshall Space Flight Center, Huntsville, AL 35812, USA}

\author[0000-0003-1533-0283]{Sergio Fabiani}
\affiliation{INAF -- IAPS, via Fosso del Cavaliere, 100, Rome, Italy I-00133}

\author[0000-0001-5717-3736]{Dawoon E. Kim}
\affiliation{INAF -- IAPS, via Fosso del Cavaliere, 100, Rome, Italy I-00133}
\affiliation{Università di Roma ``Tor Vergata'', Dipartimento di Fisica, Via della Ricerca Scientifica, 1, Rome, Italy, I-00133}
\affiliation{Universit\`a di Roma ``La Sapienza'', Dipartimento di Fisica, Piazzale Aldo Moro 2, Rome, Italy I-00185}

\author[0000-0003-3331-3794]{Fabio Muleri}
\affiliation{INAF -- IAPS, via Fosso del Cavaliere, 100, Rome, Italy I-00133}

\author[0000-0002-1868-8056]{Stephen L. O'Dell}
\affiliation{NASA Marshall Space Flight Center, Huntsville, AL 35812, USA}

\author[0000-0003-1548-1524]{Brian D. Ramsey}
\affiliation{NASA Marshall Space Flight Center, Huntsville, AL 35812, USA}

\author[0000-0002-0282-6952]{Alda Rubini}
\affiliation{INAF -- IAPS, via Fosso del Cavaliere, 100, Rome, Italy I-00133}

\author[0000-0001-5676-6214]{Carmelo Sgrò}
\affiliation{INFN-Pisa, Largo B. Pontecorvo 3, Pisa, Italy, I-56127}

\author[0000-0002-8665-0105]{Stefano Silvestri}
\affiliation{Universit\`a di Pisa, Dipartimento di Fisica Enrico Fermi, Largo B. Pontecorvo 3, Pisa, Italy, I-56127}
\affiliation{INFN-Pisa, Largo B. Pontecorvo 3, Pisa, Italy, I-56127}

\author[0000-0002-9443-6774]{Allyn F. Tennant}
\affiliation{NASA Marshall Space Flight Center, Huntsville, AL 35812, USA}

\author[0000-0002-5270-4240]{Martin C. Weisskopf}
\affiliation{NASA Marshall Space Flight Center, Huntsville, AL 35812, USA}




\begin{abstract}

IXPE (Imaging X-ray Polarimetry Explorer) is a Small Explorer mission by NASA and ASI, launched on December 9$^{th}$ 2021, dedicated to investigating X-ray polarimetry allowing angular-, time- and energy-resolved observations in the 2--8 keV energy band. IXPE is in Science Observation phase since January 2022; it comprises of three identical telescopes with grazing-incidence mirrors, each one having in the focal plane a Gas Pixel Detector (GPD). In this paper, we present a possible guideline to obtain an optimal background selection in polarimetric analysis, and a rejection strategy to remove instrumental background. This work is based on the analysis of IXPE observations, aiming to improve as much as possible the polarimetric sensitivity. In particular, the developed strategies have been applied ``as a case study'' to the IXPE observation of the 4U 0142+61 magnetar.

\end{abstract}



\section{Introduction}

In X-ray astronomy, since the first observation of a source outside the Solar system --- Sco X-1 \citep{giacconi} --- the background has had an important role, in some instances limiting the exploitation of science data. During this first detection, indeed,  two other sources of counts were observed: a diffuse emission coming from the whole sky, and an instrumental background.
After sounding rockets, X-ray experiments started to be flown on satellites, a few examples were Uhuru \citep{uhuru}, OSO-8 -- which had a polarimeter on board -- \citep{oso8}, and HEAO1 \citep{heao1}. These had onboard X-rays detectors using collimators to achieve some degree of angular resolution, but their sensitivity could be increased only by a larger collecting area, which implies a higher instrumental background. Because of this, the more advanced experiments of this kind were carefully designed to minimize background intensity and/or maximize background reproducibility. A good example of those advanced collimated detectors was the PDS experiment \citep{pds} on board BeppoSAX \citep{bepposax}: source observation was achieved by using two rocking collimators --- alternatively pointing the source and the background --- so obtaining a continuous background monitoring to control systematics associated to the single collimator and to the background field. Despite half of the collecting area being used for background monitoring, this strategy was quite successful because systematic errors in the background were typically $<1$\%. This allowed for an unprecedented exploration of the spectral properties of several extra-galactic sources.

Moreover, Einstein \citep{Einstein}, adopting X-ray mirrors, changed the picture considerably: telescopes, concentrating photons on a small spot, significantly increased the ratio between source and background counting rates for point-like sources. As long as the angular extent of the source is smaller than the Point Spread Function (PSF), the better the angular resolution of the telescope, the smaller the spot and the higher the ratio between source and background counting rates. In the case of extended sources to be resolved, this is obviously no longer true, but in light of the substantial advantage offered by concentrators, in these instruments, significantly less attention was given to background reduction and reproducibility. One exception to this rule was the PSPC experiment \citep{pspc} onboard ROSAT \citep{rosat}, an X-ray imaging detector featuring an anti-coincidence system. Thanks to it, the achieved sensitivity to low surface brightness emission has been unrivaled for over 25 years. Moreover, in the last decades, Charge Coupled Device (CCD) detectors have substituted gas detectors, because of the significant improvement in the spectral resolution that could be achieved. This substantial improvement was obtained at a price: the CCD detectors that have been flown are much slower than gas detectors, implying that a reduction of instrumental background through active shielding was no longer possible. This is one of the reasons why some X-ray experiments show a high background and a limited sensitivity to low surface brightness emission. Only in the last years, new technologies have led to CCD detectors featuring anti-coincidence systems like NuSTAR \citep{nustar} and Hitomi \citep{hitomi}.

Also, the orbit affects the background properties: High Earth Orbits allow long uninterrupted observing windows, but suffer from a higher and far less predictable background than Low Earth Orbits; these latter ones lead to a lower and more stable background thanks to the shielding provided by Earth's magnetic field but suffer from source occultations.

IXPE (Imaging X-ray Polarimetry Explorer) is a new NASA Small Explorer (SMEX) mission launched on 9$^{th}$ December 2021 by a Falcon 9; the mission operates at a 600 km altitude equatorial orbit and is fully dedicated to X-ray polarimetry \citep{ixpe2}. IXPE hosts three telescopes with a Gas Pixel Detector (GPD) \citep{gpd1,ixpe1,Sgro21} in the focal plane. It started its observation plan in January 2022, and during its first year, it is obtaining scientifically meaningful measurements of the X-ray polarization in several sources, allowing us to improve our knowledge of their geometry and emission mechanisms. Besides measuring X-ray polarization, the telescopes provide moderate angular resolution over a useful 9'$\times$9' field of view, moderate spectral resolution, and excellent timing. Pointing-system capabilities were enhanced to support ``dithering''--controlled and small--amplitude oscillations of the pointing direction to compensate for the low level of residual systematic.

In this paper, the IXPE detectors are briefly described to show the impact of background on them. Moreover, background handling for polarimetric data analysis is presented with particular care to its selection and rejection, aiming to reach the best achievable sensitivity. Then the 4U 0142+61 magnetar IXPE observation is analyzed to show the impact of this rejection+selection strategy on the results.

\section{The impact of background on IXPE focal plane detectors}

Before IXPE, only one significant measurement of X-ray polarization had been obtained, in 1975 by OSO-8. This mission had two graphite polarimeters onboard, which allowed it to obtain a precise polarization measurement of the X-ray flux from the Crab Nebula \citep{crab1,crab2}. Only in 2018, a new mission for X-ray polarimetry was launched: the CubeSat mission named Polarimeter Light (PolarLight). Based on the GPD, PolarLight demonstrated the potentiality of this detector, despite limitations due to the design typical of CubeSat (small area, no optics), it obtained a new measurement of X-ray polarization for the Crab Nebula \citep{feng} and Sco X-1 \citep{scox1}. 

IXPE performances allow us to improve the polarimetric sensitivity, mainly thanks to the imaging capabilities and the higher effective area obtained by putting GPDs at the focus of X-ray mirrors. To estimate the sensitivity of a polarimeter, the standard parameter to be used is the Minimum Detectable Polarization (MDP) at a confidence level of 99\%. It is defined as \citep{weisskopf2010}:
\begin{equation}
MDP_{99} =
\frac{4.29}{\mu S}\sqrt{\frac{S+B_{diff} + B_{res}}{T}},
\label{mdp}
\end{equation}
where:
\begin{eqnarray}\nonumber
S & = &  \int A(E) \epsilon(E) F(E)  \,dE \\\nonumber
B_{diff} & = & \int B_{diff}(E) \,dE \\\nonumber
B_{res} & = & \int B_{res}(E) \,dE \\\nonumber
\mu & = & \int \mu (E) A(E) \epsilon(E) F(E)  \,dE \nonumber
\end{eqnarray}
that are: $\mu$ the modulation factor, that is the response of the polarimeter to a fully polarized source, and for photoelectric polarimeters it depends on energy; $A$ the area of the source (for a focal plane instrument it is the effective area of the X-ray telescope); $\epsilon$ is the quantum efficiency of the detector; $F$ is the photon spectrum (photons$\cdot$s$^{-1} \cdot$ cm$^{-2}\cdot$ keV$^{-1}$); $B_{diff}$ is the diffused X-ray background collected by the instrument; $B_{res}$ is the residual background, that is the remaining background after all the rejection techniques (described in the following) have been applied; $T$ is the integration time (s). The smaller the $MDP_{99}$ is, the higher the reachable sensitivity is; reducing the background is crucial to reach the highest possible sensitivity. Typically, for bright point-like sources, for which it is possible to select the source region spatially, the background can be assumed to be negligible: ($S+B_{diff} + B_{res}\simeq S$), but for faint or extended sources, even the brightest ones, the background needs to be taken into account and properly handled. Indeed, the background is not polarized, and it can dilute the measured polarization degree ($P_{meas}$) of a source with respect to its true polarization ($P_{src}$):
\begin{equation}
P_{meas} = P_{src} \left(
1 + \frac{B_{res}+B_{diff}}{S}
\right)^{-1}.
\label{dilution}
\end{equation}

From equations \ref{mdp} and \ref{dilution}, the need to reduce the ratio between background and source as much as possible with a proper background-rejection approach is evident.

Typically, silicon detectors have a larger residual background than proportional counters, mainly because a suitable background rejection cannot be applied in silicon detectors. Proportional counters for background rejection can make use, for example, of the pulse-shape discrimination \citep{prop_counters} or of anti-coincidence systems that allow reaching a background rejection up to 99\%. Pulse shape analysis is a synthetic parameter that relates the extension of a track to the timing properties of the signal, and it was successfully applied also in cylindrical proportional counters, where it was used to detect the orientation of the track with respect to the anode \citep{Sanford1970} and therefore X-ray polarization, and in proportional counters with a configuration with a constant drift field and a multi-wire anode plane. 

\subsection{IXPE focal plane detectors: the GPDs}

\begin{figure}[!htb]
	\centering
	\includegraphics[width=0.5\textwidth]{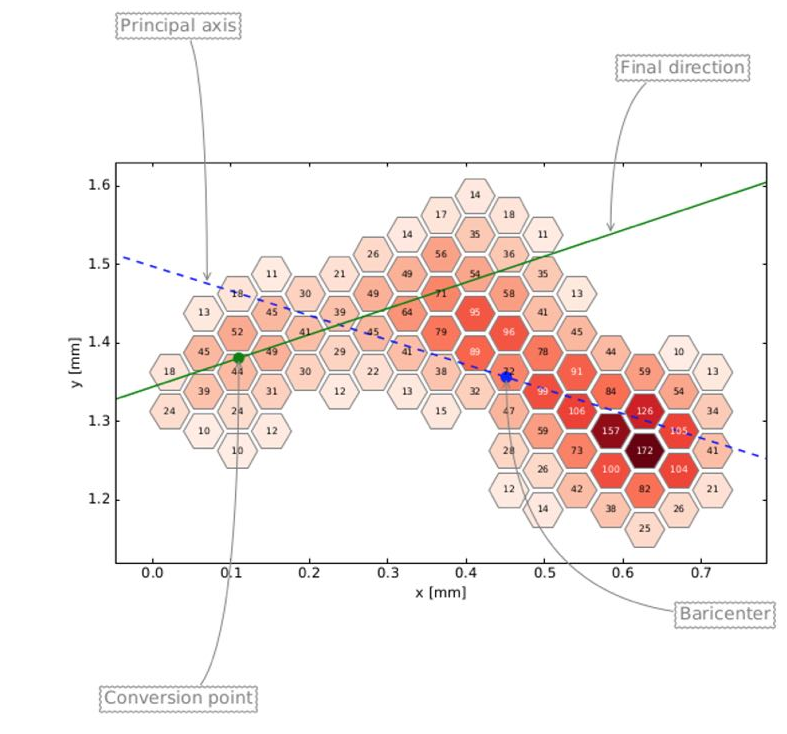}
	\caption{An example of an ionization track resulting from absorption of a 5.89 keV X-ray, as amplified by the GEM and imaged onto the GPD’s pixelated anode --- a 300$\times$352 array of hexagonal pixels at 50 $\mu$m pitch \citep{sgro17}.
	\label{fig:track}}
\end{figure}

IXPE's GPDs are able to obtain several information for each X-ray, such as energy, polarization, and time of arrival, exploiting the photoelectric effect in the gas. Summarizing, X-rays arrive through a beryllium window into the gas cell where the photon, via the photoelectric effect, ionizes an atom of the gas filling the GPD. The produced charge is driven by an electric field parallel to the optical axis toward a GEM, where the ionization is amplified before being read on the pixels of an Application Specific Integrated Circuit (ASIC). The GPD final readout is an image of the ionization track (see the example in Figure \ref{fig:track}) containing the collected charge for each pixel.
The GPD gas cell is filled with Dimethyl Ether (DME) gas at 0.8 atm, which is closed on the top by a Titanium frame and a Beryllium window, and on the bottom by the ASIC. On the sides of the gas cell are Macor spacers. The DME gas, used in the IXPE GPD, has a low diffusion coefficient and also selects the energy range, of 2--8 keV. Photoelectron tracks collected by the GPD are analyzed by a custom algorithm to extract relevant information. In this algorithm, the photoelectron track direction is determined on the basis of a two-step moment analysis that has been refined over the years \citep{Bellazzini03a,Bellazzini2003b,Sgro21,weights}. In \cite{weights}, a summary of this algorithm is described.

In the GPD, the diffused X-ray background has a very low impact on the total background due to the typical Half Energy Width (HEW) of the instrument; the residual instrumental background is, therefore, much larger. The latter one can be rejected only by using track parameters estimated in the track reconstruction algorithm. An approach based on pulse shape discrimination cannot be applied because of the lack of time distribution information for the charge during the collection on the GPD ASIC; however, the time distribution of the track is a proxy for its extension that is directly measured by the GPD. 

\subsection{Extraction of polarimetric information from IXPE data}

IXPE is providing us information on imaging, energy, time, and polarization from several categories of X-ray astronomical sources. It is sensitive to linear X-ray polarization, and it is optimized to operate in the nominal 2--8 keV energy band. 

IXPE data are distributed on HEASARC in two levels/formats: event\_l1 and event\_l2. The event\_l1 files consist of raw data containing unfiltered events, including data from in-flight calibrations, and they provide all the information about the collected tracks. The event\_l2 files are produced from event\_l1 after calibration/correction procedures. In particular, they are cleaned of in-flight calibration data and occultation/SAA passages. After this, energy calibration/equalization is applied, and the Stokes parameters are estimated from the photoelectron emission angle, $\phi_k$, for every single k-th event, as explained in \cite{Kislat}
\begin{eqnarray}
i_k&=&1 \nonumber \\
q_k&=&2\cos2\phi_k \\
u_k&=&2\sin2\phi_k,
  \nonumber
\end{eqnarray}
and a weight to optimize the polarimetric response is calculated \citep{weights}. To take into account systematic effects due to spurious modulation, this latter one is removed from the Stokes parameters on an event-by-event approach, as explained in \citep{Rankin2022}. After this latter correction, photons and their Stokes parameters are referenced to sky coordinates, removing dithering pattern and boom motion effects.

The IXPE nominal 2--8 keV energy band is constrained by the effective area of the Mirror Module Assemblies (MMA), and the efficiency of the Detector Unit (DU) placed in its focal plane \citep{ixpe1,energy}. The effective area of the MMAs has slight variations in the 2--7 keV energy band ($\sim 168$ cm$^2$ at 2.3 keV and $\sim 195$ cm$^2$ at 4.5 keV), while at 1 keV and at 8 keV it drops to $\sim 50$ cm$^2$. Also, the DU efficiency, that is the convolution of the GPD quantum efficiency and the X-ray transmission of an ion-UV filter \citep{UV} mounted on top of the DU, has a maximum ($\sim20$\%) at 2 keV with a quick drop at lower energies ($\sim0.1$\% at 1 keV), and a slower decrease up to 8 keV where it reaches a value of $\sim1$\%. IXPE telescopes are numbered from 1 to 3 to be identified, the same nomenclature is applied to MMAs, and DUs included in them.

In the following sections, a background optimal selection approach is shown on the basis of studies of the background induced polarimetric signal as estimated from event\_l2 files. To identify a possible background rejection approach, we started from the event\_l1 files containing several parameters for each track, then the best set of them is identified to obtain a set of constraints to tag background events. Each photon in the event\_l2 data is also present in the event\_l1, and they can be identified by using the TIME information. Thus, we can select the background events in event\_l1 files, then they can be removed from the TIME identifier from the event\_l2 photon list to obtain a background rejected data set to be characterized, as shown in the following. In the end, to show the effects on one of the IXPE's observations, this approach is applied to data analysis of 4U 014+61 magnetar.

\section{Background selection}\label{sec:selection}

In IXPE, the observing strategy is ''point-and-stare``, which means that one source at a time is observed, best-centered with respect to the sensitive area of the three detectors. In the case of point-like sources, the source can be spatially selected in a circular region with a radius that is at least equal to the IXPE PSF (HEW$\simeq$ 30''). Similarly, the background can be selected in a circle or an annular region outside the source region. In this section, the IXPE observation of the 4U 0142+61 magnetar, because of its high exposure ($\sim 2$ weeks) and its low flux (unabsorbed flux $\sim 7\times 10^{-11}$ erg$\cdot$s$^{-1}\cdot$cm$^{-2}$ in the 2--10 keV, \cite{Taverna2022}), is used as a typical case study to characterize the background behavior when these selections are applied. The 4U 0142+61 observation was performed from January 31 at 07:23 UTC until February 14 at 23:44 UTC and from February 25 at 04:38 UTC until February 27 at 18:46 UTC.

\begin{figure}[!htb]
	\centering
	\includegraphics[width=0.49\textwidth]{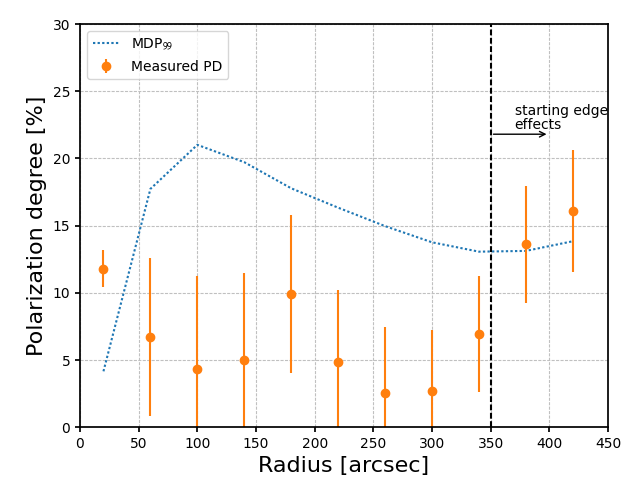}
	\includegraphics[width=0.49\textwidth]{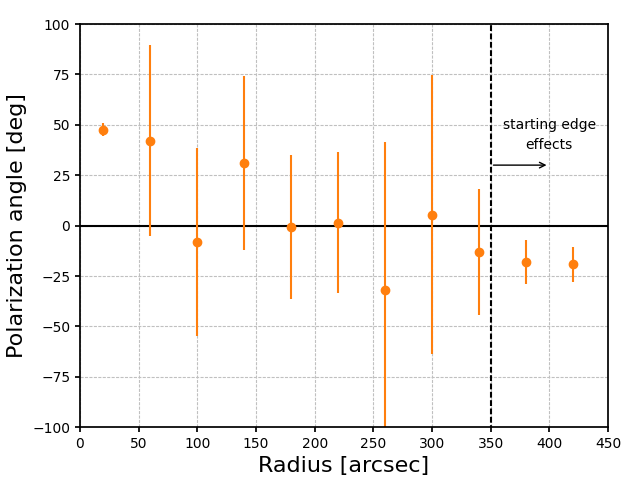}
	\caption{Polarization degree (left) and angle (right) in the IXPE nominal energy band 2--8 keV as a function of the radius for the IXPE observation of 4U 0142+61, in orange, and the corresponding MDP$_{99}$, in dotted blue line: the source is significantly polarized (first point within a circular region with radius 40''), and the background induced polarization signal (see text) is below the MDP$_{99}$ up to $\sim350$ arcsec, where it starts to be significant because of geometrical effects due to the shape of the IXPE detectors.
		\label{fig:radial}}
\end{figure}

Particles, inducing instrumental background, don't have polarization, but they can induce a modulation mimicking the X-ray polarization signal. Being not possible in IXPE to obtain particle identification, we are not able to distinguish between the X-ray polarimetric signal and the fake modulation induced by the particle background; thus, we will treat this fake signal as a background induced polarization in the following. To estimate this effect along the radial distance from the point-like source, we spatially selected the source first in a circular region with a 40'' radius, then in an annular region with a 40'' of difference between the inner and the outer radius. We increased the external radius (and consequently the internal one) up to 440''. The polarization as a function of the mean radius is shown in Figure \ref{fig:radial} for events in the IXPE nominal energy band of 2--8 keV.
The first data point is, by the fact, the source polarization in a small circular region with a 40'' radius; the source shows a significant polarization as described in the IXPE discovery paper \citep{Taverna2022}. The polarization degree reaches values compatible with zero starting from $\sim$100'', up to 300'' from the center. The polarization angle shows values compatible with the source up to $\sim$100'', from which it starts to be compatible with zero up to $\sim$350''. The regions more distant than $\sim$350'' from the center show polarization above the MDP, as shown in Figure \ref{fig:radial}. The MDP value increases because of the lower counting rate in the annular regions up to 100'', then starts to decrease due to larger regions starting to include more events up to $\sim$350'',  where edge effects begin. To have a safe annular selection of the background, it should have an inner radius of 150'', (i.e. 2.5 arcmin, which is about 5 times the IXPE HEW diameter, reducing the probability to have source events) and an outer radius of 300'' (5 arcmin) to avoid geometrical edge effects due to the detectors' shape. For the source selection, circular regions up to $\sim$100'' should include more events from the source than the background, allowing to select a region as large as possible to decrease the MDP$_{99}$ as much as possible, avoiding dilution effects due to background.

\begin{figure}[!htb]
	\centering
	\includegraphics[width=0.5\textwidth]{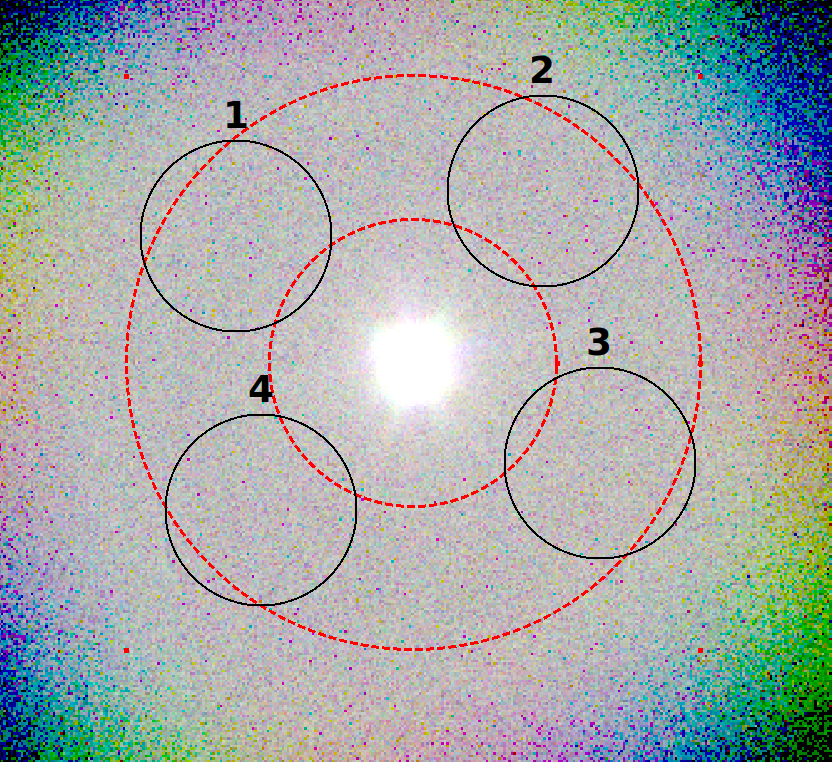}
	\caption{IXPE observation of 4U 0142+61 overimposed with four circular regions, in black, with radius 100'' and an annular region, dotted line in red, with inner radius 2.5' and outer radius 5'. These regions are used to test and compare different background selections in this analysis.
		\label{fig:regions}}
\end{figure}

\begin{figure}[!htb]
	\centering
	\includegraphics[width=0.49\textwidth]{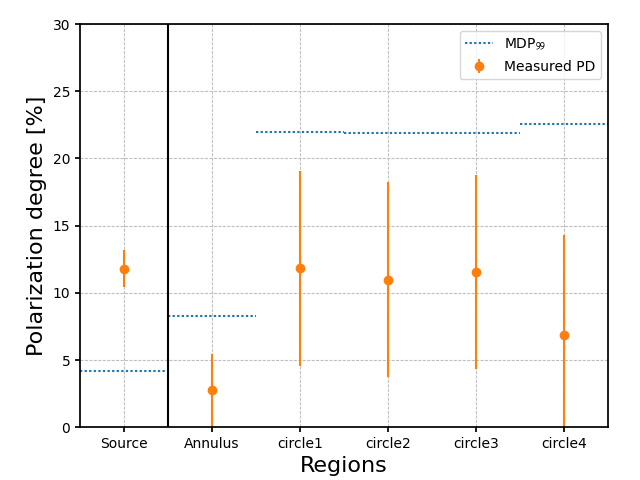}
	\includegraphics[width=0.49\textwidth]{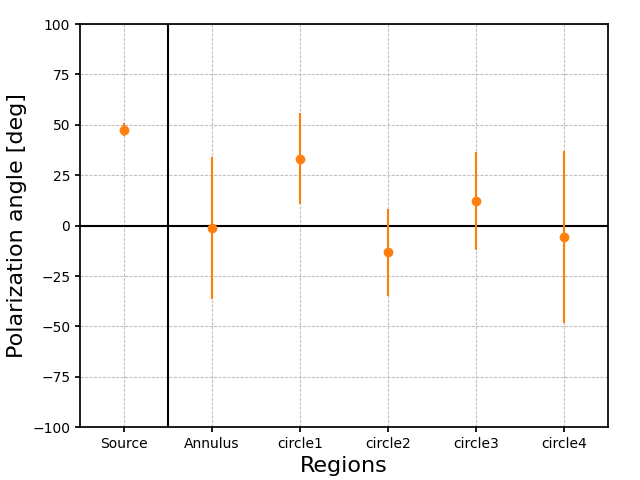}
	\caption{Induced polarization degree (left) and angle (right) from the background as measured in the different selection regions shown in Figure \ref{fig:regions}, compared with the source polarization: the polarization degree induced by the background is better estimated in the annular region.
		\label{fig:circle}}
\end{figure}

A second possible choice in the background selection is the use of circular regions, but from the previous considerations, to avoid contribution from the source and/or from detector edge effects, such circular regions can have a maximum radius of $\sim$100''. In Figure \ref{fig:regions}, the annular region obtained with the previous prescriptions is compared with 4 possible circular selections for which the background induced polarization is estimated, and in Figure \ref{fig:circle}, the background induced polarization degree arising from these regions is compared: a circular selection has a larger uncertainty on the background estimation with respect to the annular region; moreover the mean values, despite being below the MDP$_{99}$, are typically higher than the one obtained from the annular selection, and this means that in background subtraction, we are associating to the background a higher induced polarization with higher uncertainties. 

In conclusion, the best selection for the background from a polarimetric point of view is obtained by using an annular region with an inner radius of 2.5' and an outer radius of 5'. This is somewhat expected because annular regions exploit the azimuthal averaging of possible modulated components of the background.

\section{Background rejection}

IXPE is designed in order to avoid that X-ray photons in the nominal range of energy can arrive at the sensitive parts of the detectors only through the optics. Photons from the diffuse cosmic ray background not reflected by the optics are stopped by the opaque structures: the mirror block, the long collimators and, aiming to complete the coverage, dedicated skirts around the mirrors. Therefore we can assume that the photons detected in the IXPE nominal energy range are from the field of view and belong either to the source of interest or to the surrounding X-Ray background.

Background rejection methods are based on differences between the signal of interest, in our case X-rays, and other detected signals produced from other particles. In the case of IXPE GPDs, the possibility to disentangle the events due to ``good'' X-ray photons, from those constituting the background, is based on the different morphology of the recorded tracks derived from processes following the different interactions, i.e. ``good'' photons are detected by the photoelectrons. These interactions are always detected in the GPDs in the form of ionization tracks. In most cases, such as electrons, positrons, muons, or protons, the sources of background arrive from outside or are produced in the satellite's passive structures. Their energy is high, and consequently, their linear ionization loss is lower than that of photoelectrons. They release energy in the IXPE range because they lose part of their energy outside the active volume, where, instead, they induce tracks typically more extended, straight, and with a lower charge density than photoelectrons. However, the detectors are not imaging the ionization tracks in 3 dimensions, but only their projection on the sensing plane. When a particle crosses the detector at a small angle from the perpendicular, the detected track can be as long as that of a photoelectron but more straight and with constant ionization. A last complication is due to the triggering logic. In order to contain the dead time, the GPD ASIC only routes to the output the content of the Region of Interest that, because of pixels clustering and of its fixed size, can leave out the content of some pixels.

Following these considerations, we started a systematic study of possible selection criteria to separate good tracks from background ones. We aim, in this first stage, to a simple and robust method. In this spirit, given the extreme variability of the shape of good tracks, we haven't yet applied any analysis based on the track shape.

\subsection{IXPE pre-launch background study}

A preliminary study of IXPE background was performed using a Geant4 simulator in \cite{fei}: in this paper, the origin and the phenomenology of the background were briefly discussed, and the relative fluxes have been estimated (see Table \ref{tab:xie} for a summary). The main background contribution in the IXPE nominal energy band arises from the primary protons, primary $\alpha$s, and secondary cosmic rays, while $\gamma$-origin background (including cosmic X-ray and albedo $\gamma$) is not significant. The total counting rate for the instrumental background results to be 6.31$\times$10$^{-4}$ counts/sec/arcmin$^2$.

\begin{table}[!htb]
	\centering
	\begin{tabular}{c|c}
		\hline
		Component & Expected rate in 2–8 keV (counts/s) \\ \hline \hline
		Cosmic X-ray & 1.73$\times 10^{-3}$ \\
		Albedo Gamma & 1.24$\times 10^{-3}$ \\
		Albedo Neutron	& 2.97$\times 10^{-4}$ \\
		Primary Proton	& 3.16$\times 10^{-2}$ \\
		Primary Electron & 2.39$\times 10^{-4}$ \\
		Primary Positron & 1.91$\times 10^{-5}$ \\
		Primary Alpha & 1.09$\times 10^{-2}$ \\
		Secondary Proton & 1.41$\times 10^{-2}$ \\
		Secondary Electron & 1.11$\times 10^{-2}$ \\
		Secondary Positron & 3.36$\times 10^{-2}$ \\\hline
		Total & 1.05$\times 10^{-1}$ \\ \hline
	\end{tabular}
	\caption{Count rates from all the background components estimated for IXPE in \cite{fei}.}
	\label{tab:xie}
\end{table}

In \cite{fei}, a parametrization of background tracks to identify and reject them with respect to X-rays tracks induced by a source has been proposed at the cost of also removing a relatively high fraction of good events ($\simeq 20$\%). The main track properties estimated in the IXPE track reconstruction algorithm, see also \cite{weights}, that were considered in \cite{fei} are:
\begin{enumerate}
	\item Pulse Invariant (PI): gain equalized sum of the charge collected in the track, which is proportional to the detected energy.
	\item Track size: number of ASIC pixels above the threshold in the main cluster, that is the largest group of contiguous pixels of the event; typically, the background events produce a larger track size than photoelectrons at the same energy.
	\item Skewness: the third standardized moment, refers to the asymmetry of the energy distribution in the track along the major axis; typically, the background tracks are less skewed (straighter) than the photoelectron tracks.
	\item Elongation: defined as the ratio of the longitudinal over transverse components of the second moment of the track; typically the background tracks, being straighter, have a larger ratio.
	\item Charge density: defined as the energy (PI) divided by track size; typically, the background induced by (minimum) ionizing tracks has a smaller charge density.
	\item Cluster number: number of clusters after applying clustering algorithm in the ROI; typically X-rays have only one cluster while minimum ionizing particles (MIPs) produce multiple clusters.
	\item Border pixels: number of pixels in the track which are at the edge of the track ROI; typically background tracks intercept the border pixel.
\end{enumerate}
The resulting best-rejection approach of \cite{fei} is based on constraining a few parameters as a function of energy to select only source events, the constraints to be applied are summarized in Table \ref{tab:bkg_rej}.

\begin{table}[!htb]
	\centering
	\begin{tabular}{c|c|c|c|c}\hline
		Energy bin & Track size & Skewness & Elongation & Charge density \\ \hline \hline
		2.0 -- 3.5 keV & 26--67 & -0.383--0.383 & 1.023--1.506 & 116.702--315.914 \\
		3.5 -- 5.0 keV & 34--92 & -0.620--0.620 & 1.044--2.414 & 140.186--315.930 \\
		5.0 -- 8.0 keV & 48--156 & -1.013--1.010 & 1.112--4.386 & 134.936--386.334 \\ \hline
	\end{tabular}    \caption{Constraints on tracks parameters to be applied for background rejection as estimated in \cite{fei}.}
	\label{tab:bkg_rej}
\end{table}

This approach rejects about 75\% of background events at the cost of removing $\simeq 20$\% of the source events. This approach, based on MonteCarlo simulations, when applied to flight data produced discouraging results because of the large fraction of source events removed ($\simeq 50$\%) despite $\simeq$80\% of rejected background events. Such a different performance between MonteCarlo simulations pre-launch and flight data can be ascribed to different reasons: a not optimal MonteCarlo physics-list, different gain during flight operations with respect to the simulated ones, etc...

\subsection{A new rejection approach based on flight data}

From \cite{fei}, background events are expected to have longer tracks in IXPE GPDs, with a smaller charge collection per pixel because of a nearly constant (and minimum) energy loss and a larger probability of having active pixels in the track border. In this study, all the available IXPE track parameters were considered in order to identify the best ones to decouple background and source contributions. These parameters are not available in the IXPE final photon list, distributed in  event\_l2 files, but they can be retrieved from raw data included in the event\_l1 files, both files are distributed on HEASARC. To identify a photon in the event\_l2 files inside the event\_l1 files, it is possible to use the keyword 'TIME'.

In order to obtain a new rejection strategy, we adopted the safer approach to maximize the fraction of X-ray source acceptance avoiding a background rejection that removes a large fraction of source events. At this aim, we used Cyg X-2 data --- the brightest point-like source observed by IXPE at the time of this work with flux $\sim10^{-8}$ erg$\cdot$cm$^{-2}\cdot$s$^{-1}$ --- to chose the parameters to use and for their parametrization. This source has been observed by IXPE from 2022 May 2$^{nd}$ up to 2022 May 3$^{rd}$. Spatial selections, as described in Section \ref{sec:selection}, allowing to identify the source (central circular region with radius 60'') and the background (annular region with inner radius 2.5' and outer radius 5') were applied. 

For each track property, the background and source populations were compared, aiming to identify parameters having strongly different behaviors. We identified three promising track parameters that are presented in the following: (i) the number of pixels in the track; (ii) the energy fraction in the main cluster over the total charge collected for the single event; (iii) the number of border pixels. We discuss each one of them in the following.

The various methods of background rejection that we discuss in the next sections are suited to be applied all together or separately, or not applied depending on the source brightness, angular structure, or spectrum. Most rejection methods, finely trimmed to maximize the increase of sensitivity, would imply a change of the convolution matrix, to be recomputed every time. To allow for flexible use of the rejection methods, while avoiding this cumbersome task, we try to identify rejection filters that have a negligible impact on the efficiency and do not depend on the spectrum. In other terms, filters totally compatible with the standard software of polarimetric analysis and the existing response functions.

\subsubsection{Number of pixels}

\begin{figure}[!htb]
	\includegraphics[width=0.49\textwidth]{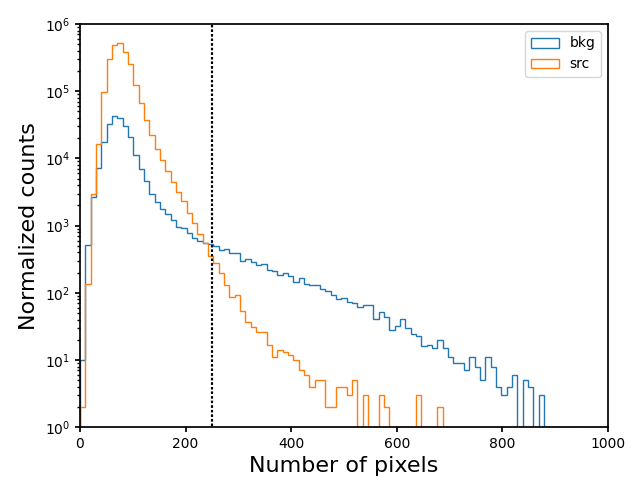}    
	\includegraphics[width=0.49\textwidth]{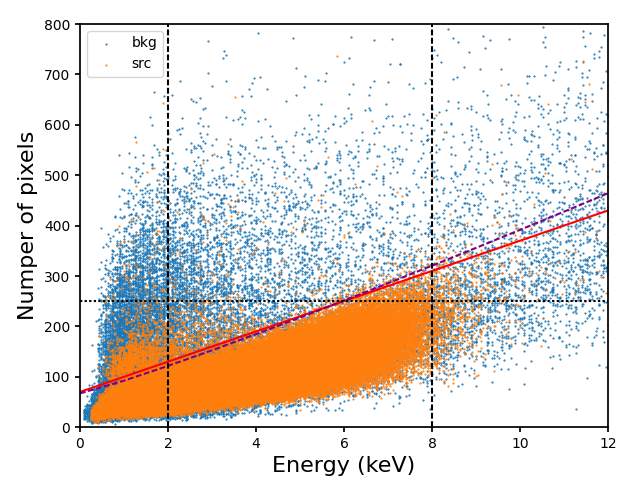}
	\caption{Number of pixels distributions for the source (orange) and the background (light blue) events from Cyg X-2 in IXPE DU1, as a representative case. Left: comparison between the two distributions in the IXPE nominal energy band, the black dotted line shows the threshold (250 pixels) above which background is dominating on source events in both plots. Right: scatter plot showing the number of pixels as a function of the energy for the source (orange) and the background (light blue) populations for $2 \times 10^5$ events from Cyg X-2. Vertical black dashed lines show the IXPE nominal energy band. The threshold values on the number of pixels as a function of energy following function \ref{eq:numpix} with best-fit parameters from Figure \ref{fig:cut_fit_pix} (purple dashed line) and the approximated linear behavior (red line) are very similar in the 2--8 keV energy band.}
	\label{fig:numpix_scat}
\end{figure}

The number of pixels is a parameter growing with energy for X-rays, this is due to the fact that typically tracks at higher energy are bigger than the ones at lower energy. Background events produce bigger tracks than X-rays when the same PI is measured. In Figure \ref{fig:numpix_scat}--left, it is possible to observe that for events in the IXPE nominal 2--8 keV energy band, source, and background events show different behaviors: in particular for a number of pixels higher than 250 the background starts to be dominating.
In Figure \ref{fig:numpix_scat}--right, it is possible to see that the constant threshold of 250 for the number of pixels can be improved with an energy dependent threshold. Repeating the comparison of Figure \ref{fig:numpix_scat}--left at different energies within the 0--10 keV energy interval with 1 keV steps, it is possible to estimate such an energy dependent threshold value. 

\begin{figure}[!htb]
	\centering
	\includegraphics[width=0.6\textwidth]{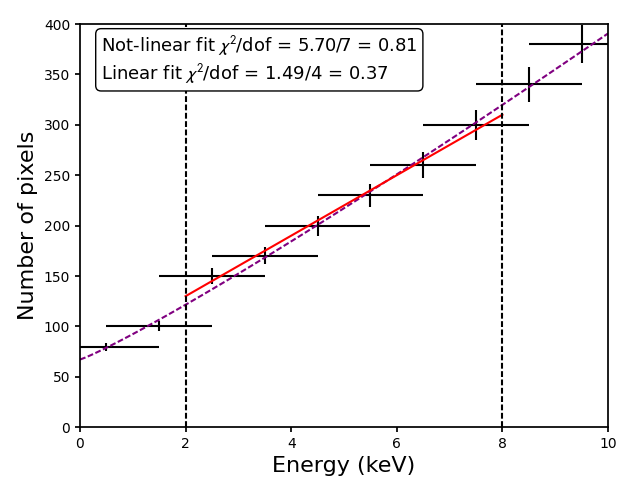}    
	\caption{Threshold on the number of pixels --- for which the source events above it are negligible --- as a function of energy (black points). The purple dashed line shows a best-fit in the 0--10 keV energy band for the function \ref{eq:numpix}, and the red line is a simplified linear function valid in the 2--8 keV IXPE nominal energy band.}
	\label{fig:cut_fit_pix}
\end{figure}

Figure \ref{fig:cut_fit_pix} shows the threshold values estimated in different energy intervals from distributions similar to the one of Figure \ref{fig:numpix_scat}--left. The threshold as a function of the energy, $E$, is well described by the function reported in \cite{muleri2010}:
\begin{equation}\label{eq:numpix}
\textrm{Number of pixels} = k_a + k_b \times \left( \frac{\textrm{E}}{\textrm{1 keV}} \right)^{k_c},
\end{equation}
with best-fit ($\chi^2/dof = 0.82$) values:
\begin{itemize}
    \item $k_a = 67.1\pm 5.6$
    \item $k_b = 25.3 \pm 4.6$
    \item $k_c = 1.108 \pm 0.084$
\end{itemize}

In the IXPE 2--8 keV nominal energy band, this equation can be simplified with a linear function ($\chi^2/dof = 0.37$):
\begin{equation}
\textrm{Number of pixels} = 70 + 30 \times E.
\end{equation}

The scatter plot of Figure \ref{fig:numpix_scat}--right shows a comparison between the different threshold functions, in the IXPE nominal 2--8 keV energy band, the two are consistent, while the fixed threshold at 250 number of pixels is not effective like these energy dependent cuts.

\subsubsection{Energy fraction}

\begin{figure}[!htb]
	\centering
	\includegraphics[width=0.49\textwidth]{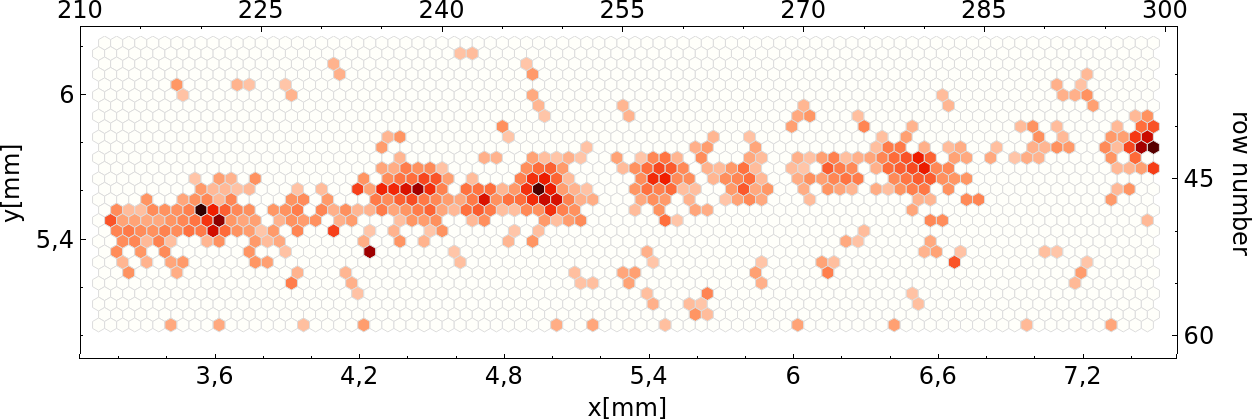}
	\includegraphics[width=0.49\textwidth]{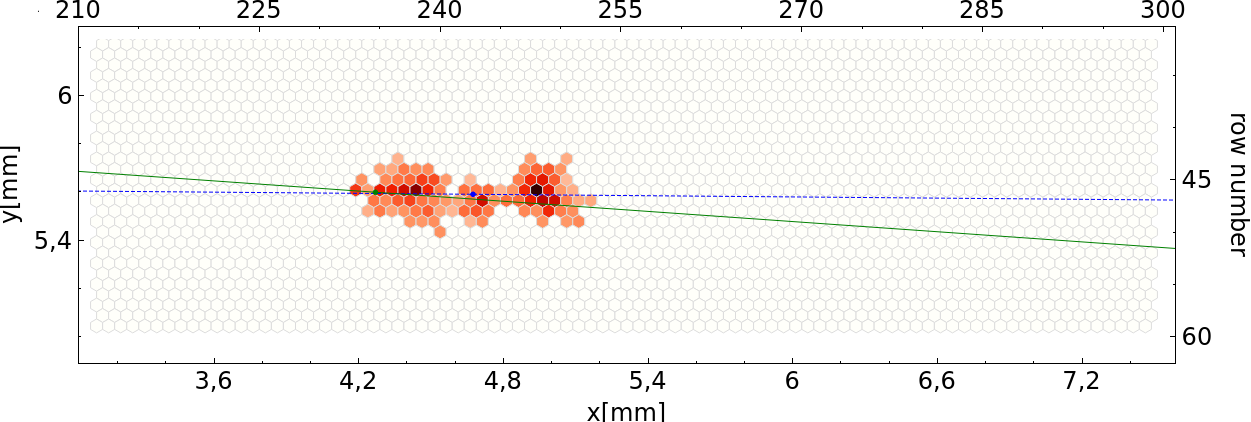}
	\caption{Example of a track with more clusters (left), where only the one reported in the right panel is considered for the track reconstruction by the IXPE algorithm.}
	\label{fig:evt_fra_track}
\end{figure}

A readout track from the GPD can have more clusters of pixels, depending on the energy density and the track threshold. Only the main cluster, the larger one, is considered for the event reconstruction \cite{Sgro21}, as shown in Figure \ref{fig:evt_fra_track}, where the raw track is represented on the left panel and the main cluster, considered for the track reconstruction, on the right one.

The parameter here considered, the energy fraction, takes into account the ratio between the energy/charge collected in the main cluster and the one collected in all the detected clusters. 
\begin{figure}[!htb]
	\includegraphics[width=0.49\textwidth]{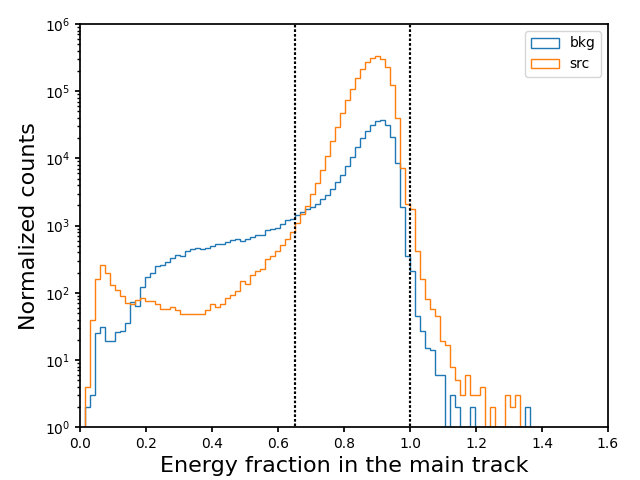}
	\includegraphics[width=0.49\textwidth]{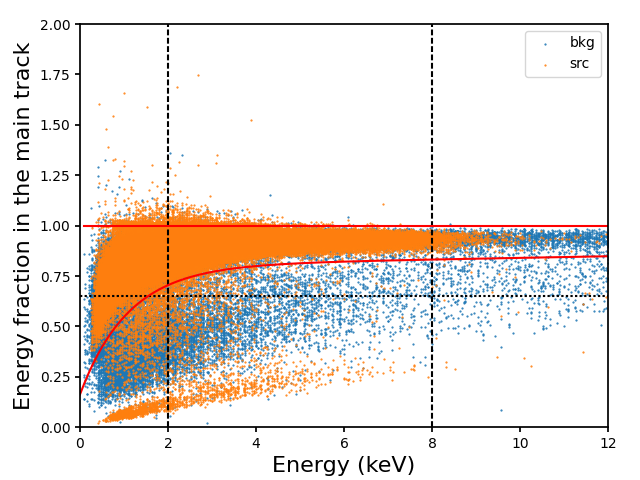}
	\caption{Energy fraction in the main cluster for source (orange) and background (light blue) events in IXPE DU1 as a representative case. Left: comparison between the two distributions in the IXPE nominal energy band, the black line shows the threshold above which background is dominating on source events in both plots. Right: scatter plots showing the energy fraction in the main cluster as a function of energy for $2 \times 10^5$ events from Cyg X-2. Vertical black dashed lines show the IXPE nominal energy band 2--8 keV. The threshold values on the energy fraction in the main cluster from the physical condition $\leq 1$ and from the energy dependent function estimated from fit (red lines) are reported. For this bright source, it is possible to identify also a population of events with a small energy fraction $\leq 0.25$ probably due to events produced by simultaneous tracks (pile-up).}
	\label{fig:efra_scat}
\end{figure}
The energy fraction definition shows that X-rays sources can have an energy fraction $\leq 1$, but from Figure \ref{fig:efra_scat}, it is possible to observe a small tail of events with a higher energy fraction, probably due to noisy pixels. These events having energy fraction $>1$ must be removed from the analysis. It is possible to observe, for this bright source, the presence of a population of events with a small energy fraction, $<0.25$. These are probably due to the detection of simultaneous tracks (pile-up). Cyg X-2 is a very bright point-like source, which means that photons are focused in a small region on the GPD sensitive area. In this particular case, it is possible to observe simultaneous detection of 2 X-rays in the track ROI, but only one will be included in the main cluster producing events having energy fraction $<0.5$ also in case of source events.

\begin{figure}[!htb]
	\centering
	\includegraphics[width=0.6\textwidth]{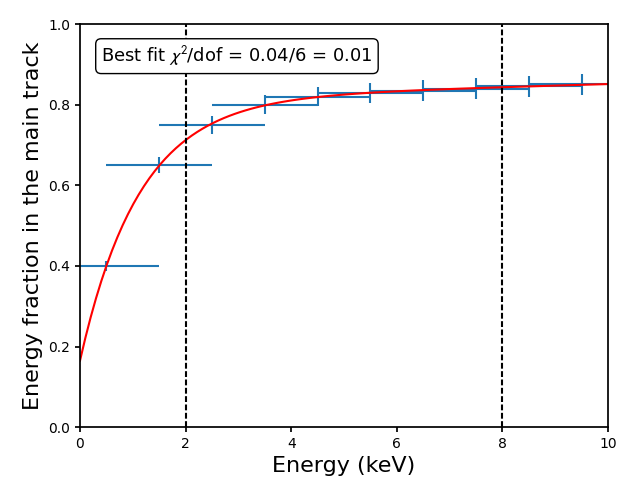}     
	\caption{Threshold on the energy fraction as a function of energy for which source events below it are negligible (light blue points). The red line shows a best-fit in the 0--10 keV energy band.}
	\label{fig:cut_fit_evt}
\end{figure}

From the point of view of the background, MIPs, characteristic of cosmic rays impinging on the detector, are expected to have longer tracks in the IXPE detectors. This means that they are able to produce more clusters along their passage through the detector, giving rise to a smaller energy fraction in the main track with respect to X-rays, like in the example of Figure \ref{fig:evt_fra_track}. From Figure \ref{fig:efra_scat}--left, it is possible to observe that a mean energy fraction in the main cluster below 0.65 guarantees that background is dominant with respect to source events. In Figure \ref{fig:efra_scat}--right, similarly to the case of the number of pixels, the energy fraction in the main cluster threshold can be optimized as a function of energy. In Figure \ref{fig:cut_fit_evt}, the threshold has been estimated at different energies within the 0--10 keV energy interval with 1 keV steps.
The energy dependent threshold on the energy fraction in the main cluster, such that background events are dominating below it, is well described by the function:
\begin{equation}
\textrm{Fraction of energy} = k_a \times [ 1 + e^{-\frac{E+k_b}{k_c}} ] + k_d \times E
\end{equation}
with best-fit ($\chi^2/dof = 0.01$) parameters:
\begin{itemize}
	\item $k_a = (0.814 \pm 0.004) \sim 0.8$
	\item $k_b =  (0.254 \pm 0.010) \textrm{ keV} \sim 0.25$ keV
	\item $k_c = 1.12 \pm 0.02 \textrm{ keV} \sim  1.1$ keV
	\item $k_d = (0.0037 \pm 0.0005) \textrm{ keV}^{-1} \sim 0.004$ keV$^{-1}$.
\end{itemize}

The scatter plot for the populations of source and background events of Figure \ref{fig:efra_scat}--right shows the threshold lines obtained from the previous fit and from the physical condition that this quantity must be $\leq 1$ compared with the fixed, energy independent, threshold from Figure \ref{fig:efra_scat}--left, showing that energy dependent threshold is more effective.

\subsubsection{Border pixels}

Background events induced by particles, as said in the previous sections, are capable of producing longer tracks and more clusters, meaning that they are also characterized by tracks with a higher probability of having pixels on the ROI edges. Moreover, border tracks are induced by events having part of the energy released outside the ROI.
\begin{figure}[!htb]
	\centering
	\includegraphics[width=0.5\textwidth]{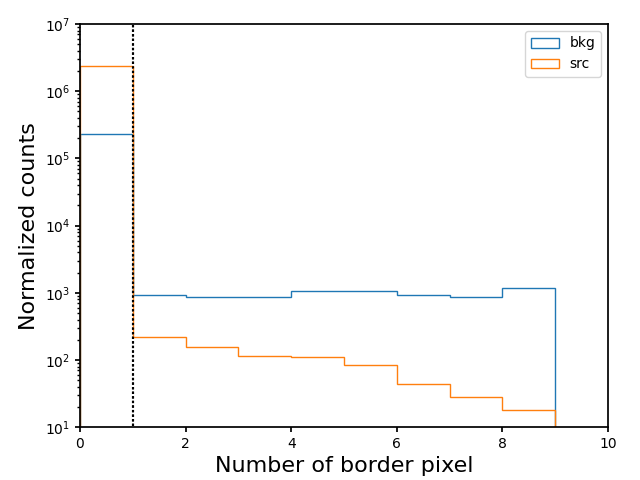}
	\caption{Comparison of source (orange) and background (light blue) distributions spatially selected by using Cyg X-2 IXPE data for the number of pixels in the border. The black dotted line shows the threshold value above which the background is dominating with respect to the source. See text.}
	\label{fig:comparison}
\end{figure}
In Figure \ref{fig:comparison}, the number of border pixels for events in the IXPE energy band is shown. An energy dependent behavior for this parameter has not been observed: at every energy, the distributions are similar; thus, the condition for background rejection is given by a cut for border pixels $<2$.

\subsubsection{Rejection strategy}\label{sec:rejection}

In conclusion, this analysis repeated for all the three IXPE detectors and for more observed sources, shows that source events must simultaneously satisfy the following conditions in the IXPE nominal energy band 2--8 keV:
\begin{itemize}
	\item Number of pixels $< 70 + 30 \times E$
	\item $0.8 \times [ 1 - e^{-\frac{E+0.25}{1.1}} ] + 0.004 \times E < $ Energy fraction $<1$
	\item Border pixels $<2$
\end{itemize}

In case of future IXPE energy band extensions, see e.g. \citep{energy}, the condition for the number of pixels should be changed with the one arising from the fit of function \ref{eq:numpix}.

\section{Background rejection application to flight data}

In this section, we apply our background-rejection strategy to IXPE flight data in order to validate its effectiveness and its negligible impact on the response matrices. The fraction of events removed in the background region will be estimated, for point-like sources observed by IXPE, together with the counting rate in the background annular region. With the aim to estimate the effect on the response matrices from this rejection approach, we applied it to in-flight calibration data and to images of extended sources.

\subsection{Background rejected fraction}

The rejection efficiency has been estimated for all the point-like sources observed by IXPE up to August 2022. In particular, the observed sources are reported in Table \ref{tab:sources}. Some of them have been observed in different times and states --- as Cen X-3, Mrk 421, ... --- in these cases in Table \ref{tab:sources} the different observations are analyzed separately, and they correspond to different raws of counting rates.

\begin{table}[!htb]
	\centering
	\begin{tabular}{c|ccc|c}
		\textbf{Source name} & \textbf{Rate (cps)} & & \textbf{Source name} & \textbf{Rate (cps)} \\\cline{1-2} \cline{4-5}
		Cen X-3    & 0.14 & & Mrk 421    & 0.32 \\ 
		& 4.3  & & & 0.59 \\ \cline{1-2}
		4U 0142+61 & 0.26 & & & 1.0 \\ \cline{1-2} \cline{4-5}
		Cen A      & 0.092 && BL Lac     & 0.024\\ \cline{1-2}
		Her X-1    & 1.20  & & & 0.043 \\ \cline{1-2} \cline{4-5}
		Mrk 501    & 0.27 & & Cyg X-1    & 12 \\
		& 0.49 & && 14 \\ \cline{4-5}
		& 0.26 & & 3C 454.3    & 0.013 \\ \cline{1-2} \cline{4-5}
		4U 1626-67 & 0.34 & & 3C 273     & 0.089 \\ \cline{1-2} \cline{4-5}
		GS 1826-238 & 9.2 & & 3C 279     & 0.020 \\\cline{1-2} \cline{4-5}
		S5 0716+714 & 0.0062 & & GX 301-2   & 0.42 \\\cline{1-2} \cline{4-5}
		Vela X-1   & 0.41 && \textbf{Background} & $\sim$0.003\\\cline{1-2} \cline{4-5}
		Cyg X-2    & 15  \\\cline{1-2} 
	\end{tabular}
	\caption{List of point-like sources observed by IXPE that have been considered in this analysis with their counting rate compared to the mean estimated background counting rate. Source counting rates are estimated per DU and in a central circular region with a radius 1'; the background value has been normalized for the same area.}
	\label{tab:sources}
\end{table}

\begin{figure}[!htb]
	\centering
	\includegraphics[width=0.47\textwidth]{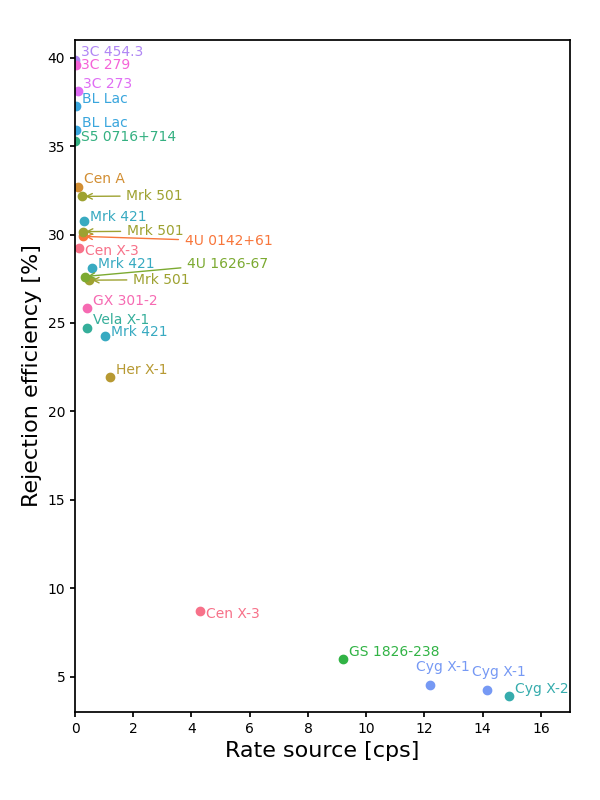}
	\caption{Fraction of background events rejected as a function of the source counting rate.}
	\label{fig:rej_eff}
\end{figure}

\begin{figure}[!htb]
	\centering
	\includegraphics[width=0.49\textwidth]{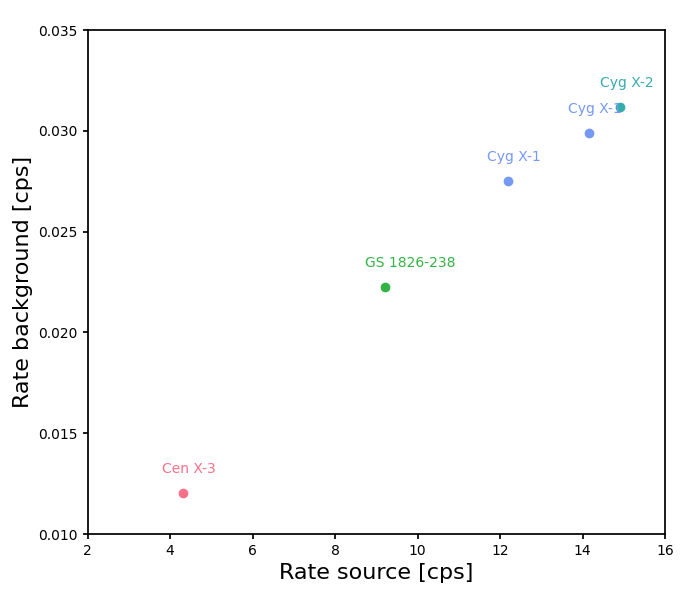}     
	\includegraphics[width=0.49\textwidth]{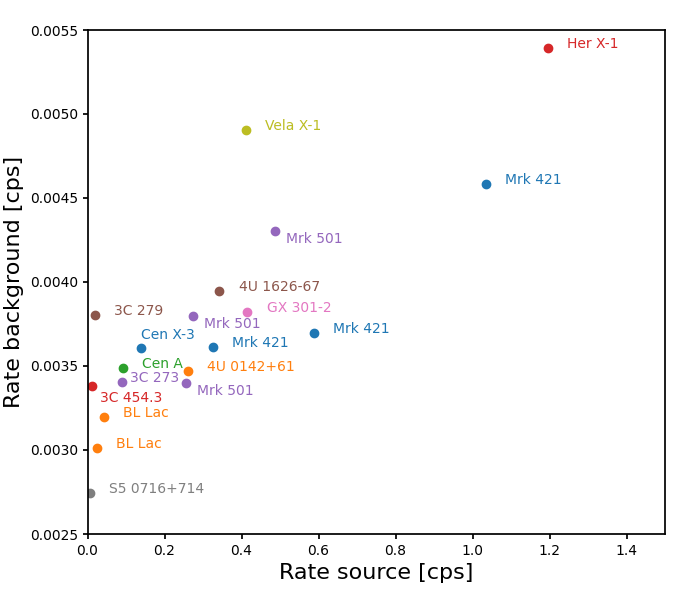}     
	\caption{Rate in the annular region where the background is estimated as a function of the source counting rate, the right panel shows a zoom for sources with lower counting rates. Both counting rates are normalized in a $\pi$ arcmin$^2$ spatial region.}
	\label{fig:rates}
\end{figure}

The rejection approach has been applied to the background region of all these sources. The background is selected as in Section \ref{sec:selection}. In Figure \ref{fig:rej_eff}, the rejected fraction of events as a function of the source rate, extracted in a circular region with radius 1', is shown.
Bright sources (inducing more than 2 cps/arcmin$^2$) show a slightly different behavior with respect to fainter ones. This is due to the fact that these bright sources, because of the IXPE PSF, contaminate the background region: this is visible from Figure \ref{fig:rates}, where the rate in the background region is shown as a function of the rate in the  source region. These plots show that background events are overwhelmed by source events also in the outer region, making rejection/subtraction of events not effective. Therefore, the background is well negligible for bright sources in IXPE.

The IXPE background is estimated, considering the faintest sources, to range from $(8.74\pm 0.18)\times 10^{-4}$ counts per second per arcmin$^2$ for S5 0716+714 (RA 07h 21m 53.45s Dec. 71 20' 36'') up to $(12.10\pm0.37)\times10^{-4}$ cps/arcmin$^2$ for 3C 279 (RA 12h 56m 11.1s Dec. -05 47' 22''). These values include both particle and X-ray background, and they are 1.3-1.9 times higher than the one estimated from simulations in \cite{fei}, where the Cosmic X-Ray Background arriving through the optics --- negligible with respect to particle background --- was not included in the simulator. When higher values of the counting rate in the background region are observed, these events are mainly due to the observed source. As an example, if one estimates for Cyg X-1, the polarization in the central circular region obtain $(3.81 \pm 0.31)$\% with polarization angle $(46.6\pm2.0)^\circ$, while the polarization in the background outer region is estimated to be $(4.0 \pm 1.7)$\% with angle $(30\pm14)^\circ$. The difference between the source and the background region is $\Delta$PD$ = -(2.0\pm1.7)\%$ for polarization degree and $\Delta$PA$ = (17\pm 14)^\circ$ for the polarization angle, both compatible with a null difference, confirming that source events are also dominating in the annular region where the background is selected.

\begin{figure}[!b]
	\centering
	\includegraphics[width=0.8\textwidth]{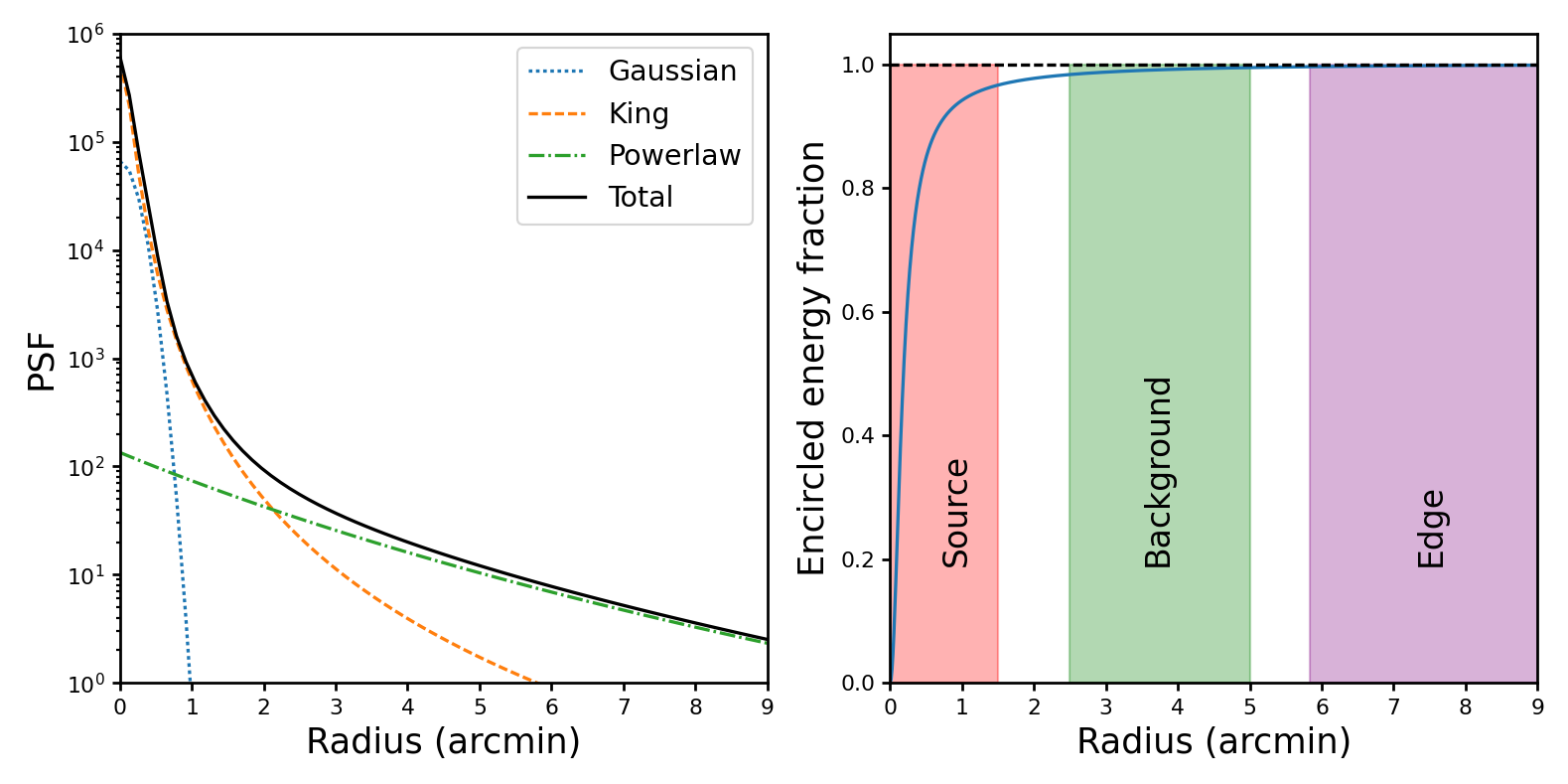} 
	\caption{Example of PSF function as estimated during IXPE telescope ground calibrations (left). And the three different regions of selection.}
	\label{fig:psf}
\end{figure}

To further confirm this effect, during the IXPE ground calibration, one telescope was calibrated, and the PSF was estimated to be composed by a gaussian plus a King function plus a power-law term (see Figure \ref{fig:psf}).
From this mean estimate obtained for the spare telescope, having the best PSF, similar to IXPE telescope 1, $\sim$97\% of events is included in a circle having radius 1.5' and $\sim$1.1\% of events is included in the background annular region. This means that for Cyg X-2, having a rate in DU1 of 15.7 cps in the source region, we expect in the background annular region $\sim 0.2$ cps.

\subsection{Effects of rejection on calibration data}

IXPE adopts a point and stare observing strategy, so that when the source cannot be observed because of Earth occultation, one detector at a time is calibrated by using onboard calibration sources \citep{fcw,monitoring}. The rejection approach proposed here has been applied to data from in-flight calibration with the unpolarized sources called Cal C, a $^{55}$Fe source having a peak at 5.89 keV, and Cal D, a $^{55}$Fe source illuminating a Silicon target producing a peak at 1.7 keV.

\begin{figure}[!b]
	\centering
	\includegraphics[width=0.32\textwidth]{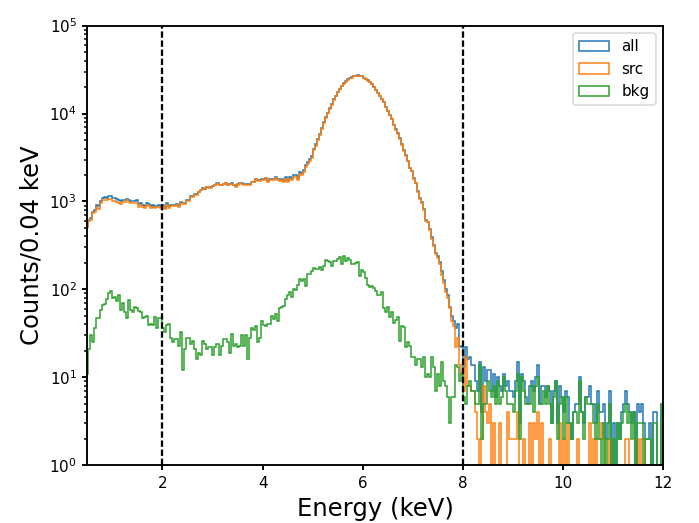}     
	\includegraphics[width=0.32\textwidth]{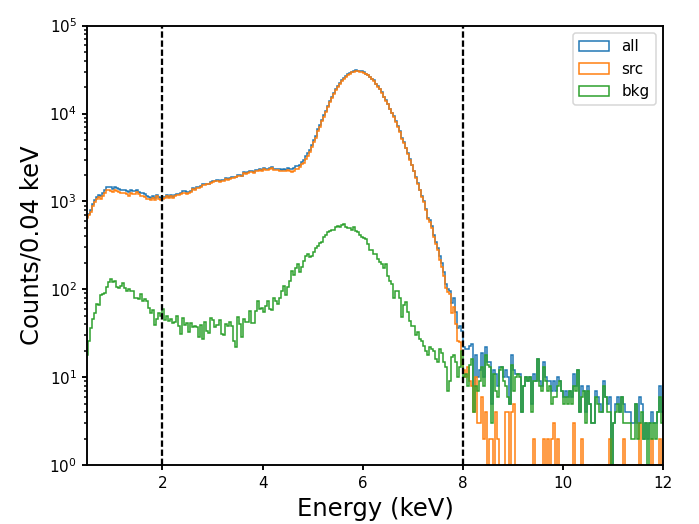}     
	\includegraphics[width=0.32\textwidth]{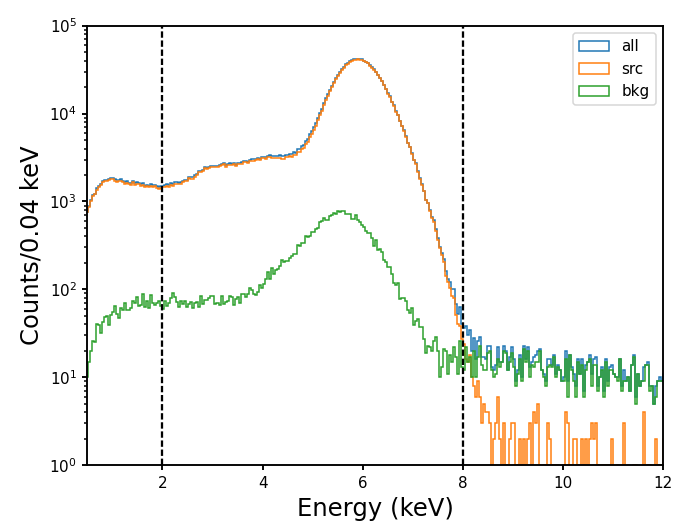}     
	\caption{Energy spectrum from Cal C in the three detectors of IXPE, from left to right: DU1, DU2, DU3. In each plot, the spectrum without rejection (blue line) with label all is compared with the one of events tagged as source (src) after rejection (orange line) and the rejected events tagged background (bkg, green line).}
	\label{fig:calc}
\end{figure}
In Figure \ref{fig:calc}, the energy spectra observed by the three IXPE detectors when exposed to Cal C source are shown and compared with events identified by this rejection approach to be source or background. It is possible to estimate that in these data sets, the rejection approach removes only 1--2\% of events, and it is able to remove the high energy tail, not due to the calibration source itself.

\begin{figure}[!t]
	\centering
	\includegraphics[width=0.32\textwidth]{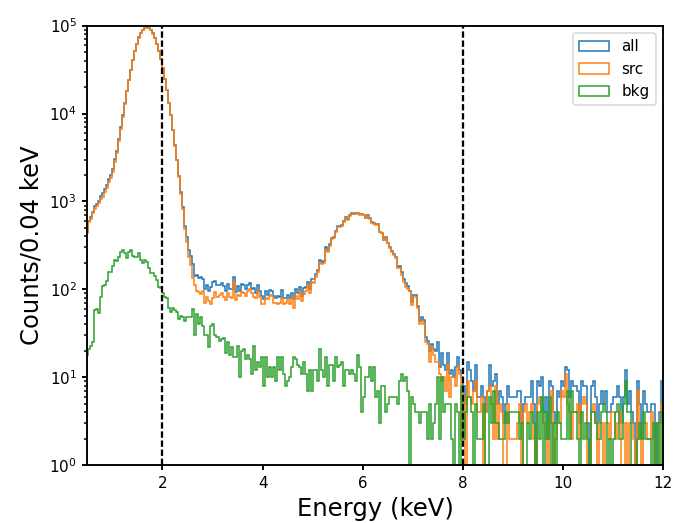}     
	\includegraphics[width=0.32\textwidth]{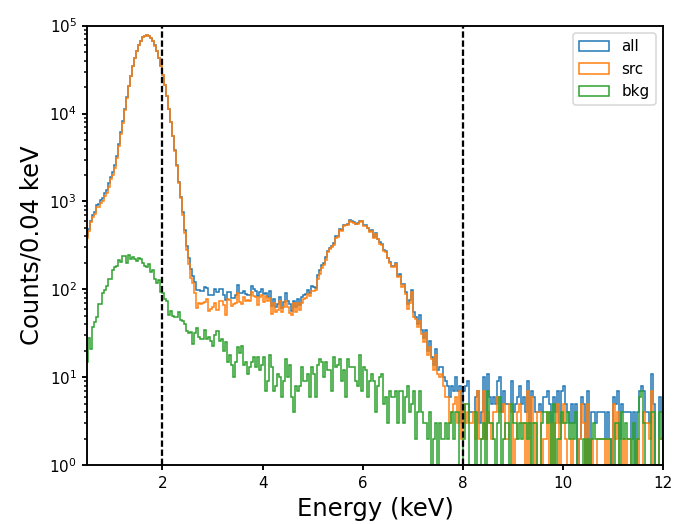}     
	\includegraphics[width=0.32\textwidth]{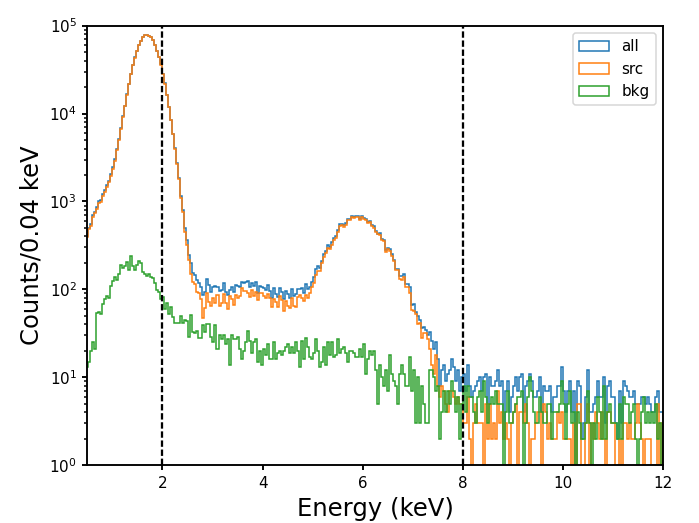}     
	\caption{Energy spectrum from Cal D in the three detectors of IXPE, from left to right: DU1, DU2 and DU3. In each plot, the spectrum without rejection (blue line) with label all is compared with the one of events tagged as source (src) after rejection (orange line) and the rejected events tagged background (bkg, green line).}
	\label{fig:cald}
\end{figure}

In the case of Cal D, whose energy peak is below the IXPE nominal energy band, the fraction of events rejected is 1--2\% in the three detectors, as shown in Figure \ref{fig:cald}.

\begin{figure}[!h]
	\centering
	\includegraphics[width=0.49\textwidth]{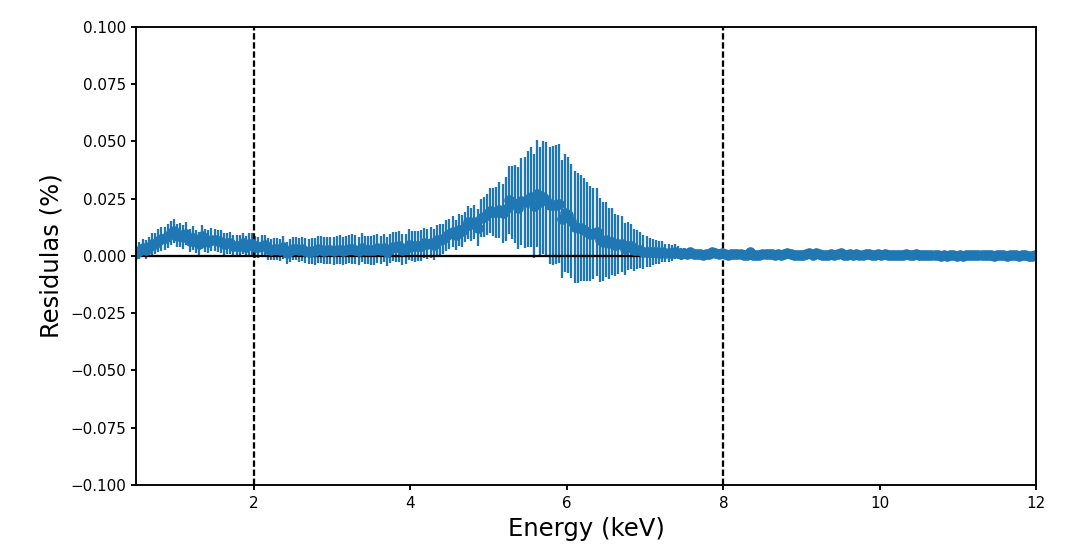}     
	\includegraphics[width=0.49\textwidth]{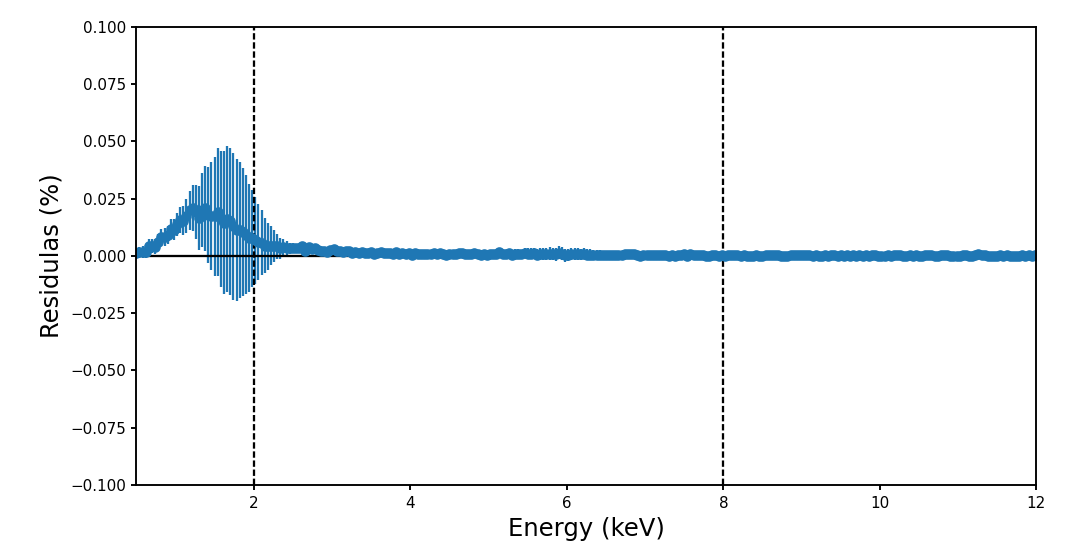}     
	\caption{Percentage residuals of the energy spectrum from Cal C (left) and Cal D (right) in one of the three detectors (DU1), as a case study, but results in the other DUs are comparable. It is possible to observe that the residuals deviate at maximum at 0.05\% from zero.}
	\label{fig:rescal}
\end{figure}

 A small spill-out of source events identified to be background is present but very small. To quantify this effect, the fraction of residual events between the energy spectra of calibration sources before and after the rejection are shown in Figure \ref{fig:rescal}: the effect is that the energy spectrum at maximum differs of 0.05\%, that is well below the uncertainty on the IXPE response functions. As a result, the effects of this rejection approach on the IXPE response matrices are well negligible. As an example, in the Cal D spectra of Figure \ref{fig:cald}, the $^{55}$Fe line is not visible in the background spectra because its source counting rate ($\sim 1000$ counts/0.04 keV) give rise to a spill-out $\leq$0.5 events/0.04 keV.

\subsection{Effects of rejection on extended sources}

\begin{figure}
	\centering
	\includegraphics[height=5.8cm]{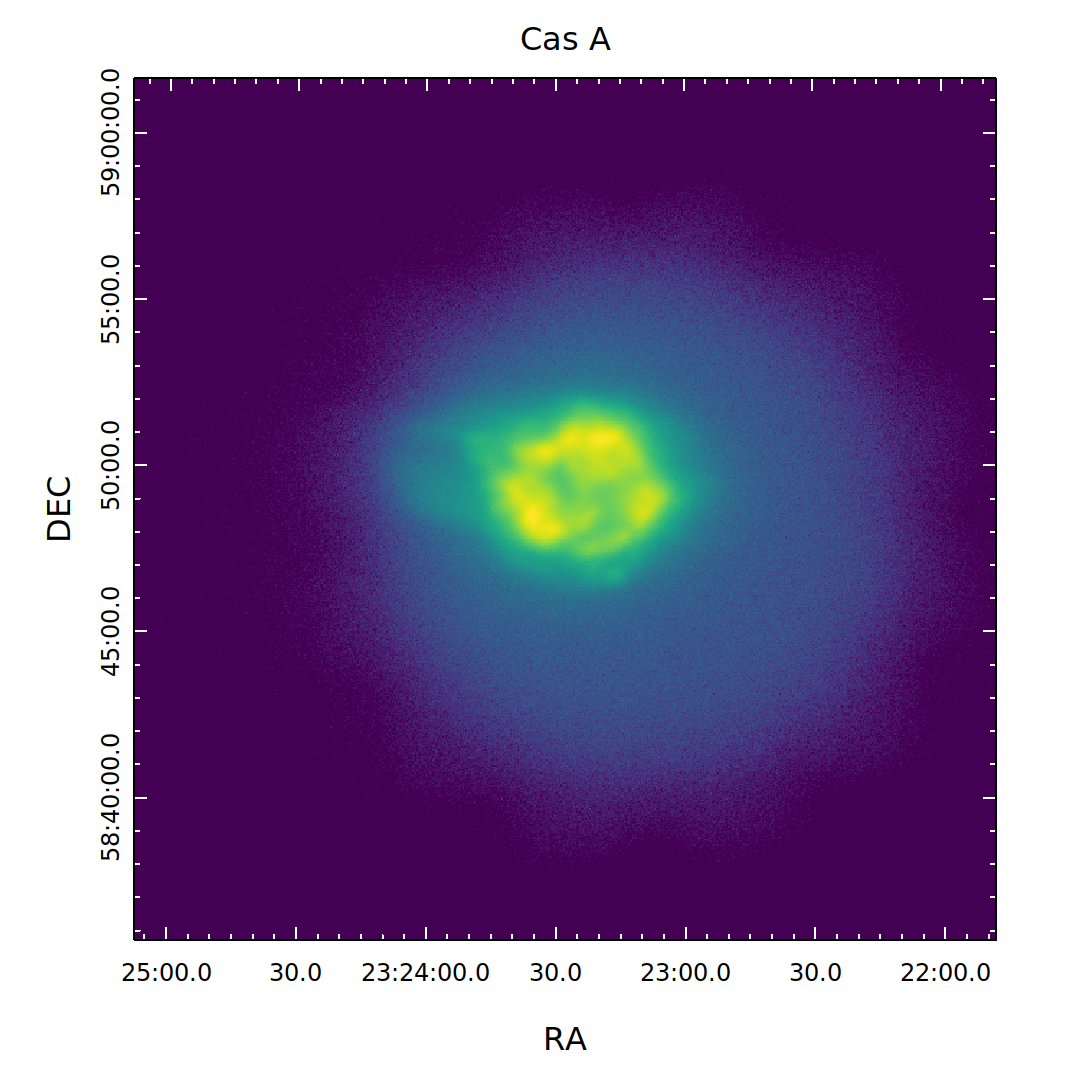}   	\includegraphics[height=5.8cm]{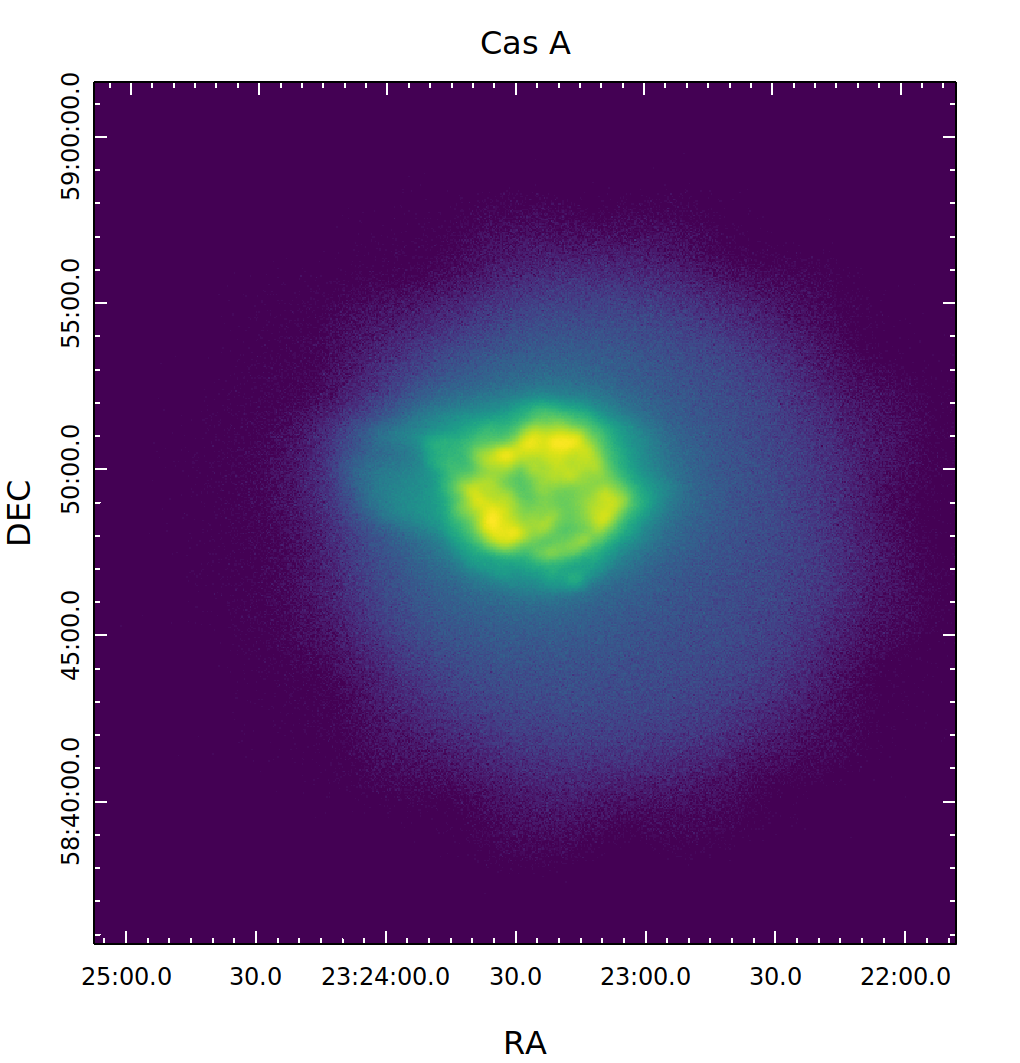}	\includegraphics[height=5.8cm]{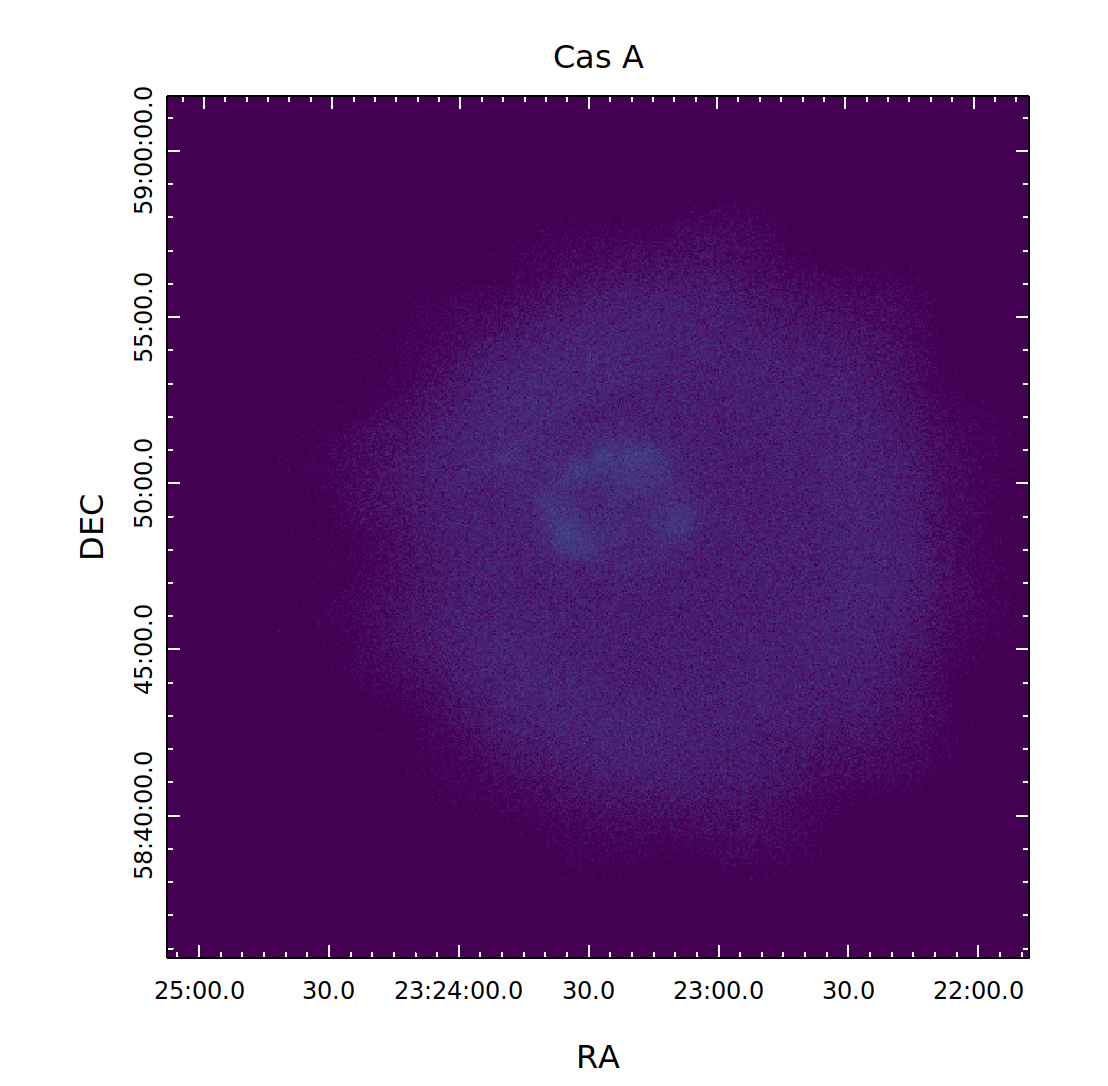}
	\caption{Cas A data images from the original data (left), after the background rejection (center), and for the rejected events (right).	The same color scale was applied.}
	\label{fig:casa}
\end{figure}

The IXPE imaging capabilities allow studying spatially resolved polarization in extended sources like Supernova remnants, Pulsar Wind Nebula, jet structures, and Galaxy clusters. The first extended source observed by IXPE is Cas A, observed from January 11 to January 29, 2022, for a total exposure time of $\sim$900 ks \citep{casa}. In the analysis of this source, reported in \cite{casa}, a preliminary version of this rejection approach was applied to allow to improve the significance of the polarimetric result. Such a preliminary approach used a fixed cut on the number of pixels, the energy fraction in the main cluster, and the border pixels. Figure \ref{fig:casa} shows the IXPE image of Cas A before (left) and after (center) the application of this new rejection approach using the selection of Section \ref{sec:rejection}. 
The rejection is not affecting imaging capabilities. Moreover, in Figure \ref{fig:casa}--right, the image of rejected events shows that they are mainly localized on the edge of the field of view of the three IXPE telescopes.

\section{4U 0142+61, a case study}

The IXPE observation of 4U 0142+61 was obtained from two segments: the first started on 2022-01-31 at 07:37:07 UTC and ended on 2022-02-14 at 23:44:12 UTC, the second started on 2022-02-25 at 04:38:09 UTC and ended on 2022-02-27 at 18:46:09 UTC, for 835719 seconds of total live-time in the three IXPE telescopes \cite{Taverna2022}. Here, the original data -- available on the IXPE archive -- have been filtered as indicated in Section \ref{sec:rejection}, and results from the data analysis before and after the background rejection are compared. 

In this analysis, after background rejection, the residual background has  also been subtracted because the source is faint. To make this, the source and background selection have been obtained following the selection prescriptions of Section \ref{sec:selection}, by using the SAOIMAGEDS9 software \citep{ds9} to produce the regions and FTOOLs included in HEASoft 6.30.1 to extract selected data.

\begin{figure}[!htb]
	\centering
	\includegraphics[width=0.5\textwidth]{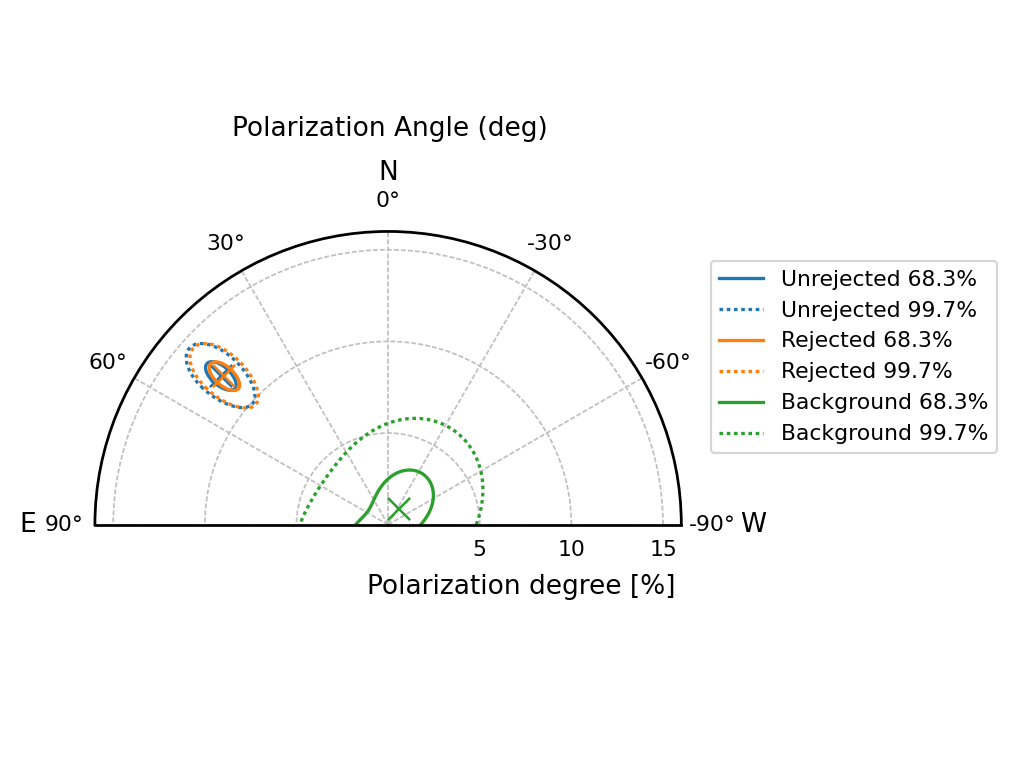}
	\vspace{-1.5cm}
	\caption{Contours enclosing regions at the confidence level of 68.3\% (solid lines) and 99.7\% (dotted lines) in case of original unrejected data (light blue), after the rejection of Section \ref{sec:rejection} (orange) and for the events rejected because identified as background (green).}
	\label{fig:pol_bkg}
\end{figure}

A preliminary overall analysis is here performed with the ixpeobssim analysis tool \citep{ixpeobssim}, developed for IXPE, considering events in the whole IXPE 2--8 keV nominal energy band. The polarization before and after the background rejection is reported in Figure \ref{fig:pol_bkg} and compared with the polarization estimated for the events removed from the rejection because tagged as background, which is well compatible with zero; polarization in the original data-set results to be PD = $(12.2\pm0.8)$\% with PA = $(48 \pm 2)^\circ$, well compatible with the value obtained after background rejection PD = $(12.1\pm0.8)$\% with PA = $(48 \pm 2)^\circ$. 

A spectral analysis was also carried out by using \textsc{xspec} \citep{xspec} version 12.12.1 released with HEASoft 6.30.1, which includes polarimetric models. Following the analysis reported in \cite{Taverna2022,Rea2007}, the standard 4U 0142+61 spectral model: \textsc{TBabs*(BBody+PowerLaw)} has been used. The interstellar absorption was taken into account by using the \textsc{TBabs} with abundances from \citep{tbabs} fixing the column density value $n_H = 0.57 \times 10^{22}$ cm$^{-2}$ \citep{nh}. The IXPE spectra from the three telescopes have been simultaneously fitted, including a cross-normalization factor for the second and the third telescope. 

\begin{figure}[!htb]
	\centering
	\includegraphics[width=\textwidth]{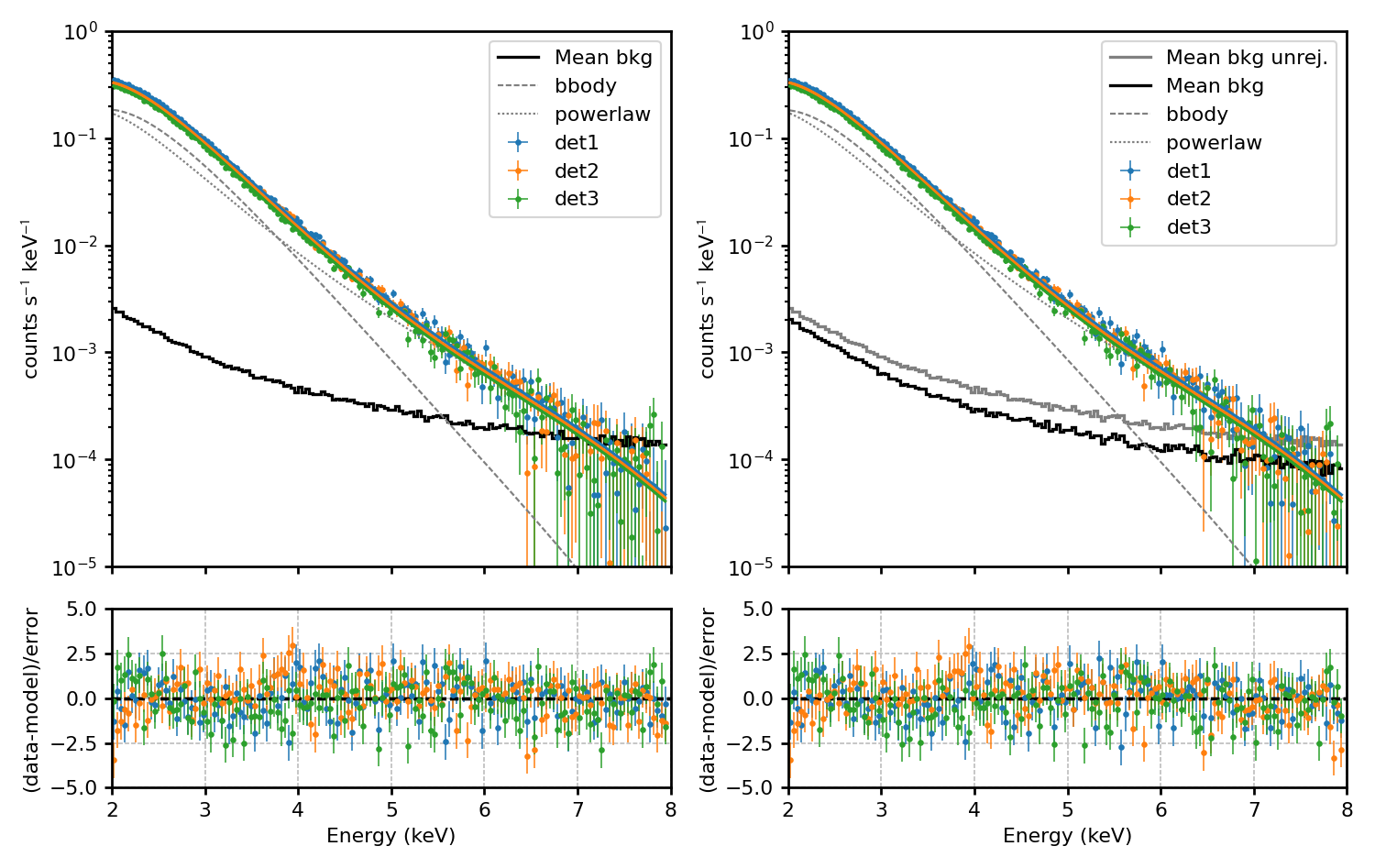}
	\caption{4U 0142+61 data fitted with \textsc{xspec} before (left) and after (right) the background rejection; the considered spectral model is \textsc{TBabs*(BBody+PowerLaw)}. At the bottom, residuals with respect to the best-fit model are shown. On the right panel, the background before and after rejection are compared, and it is possible to appreciate its reduction after the rejection procedure.}
	\label{fig:fit_xspec}
\end{figure}

Figure \ref{fig:fit_xspec} shows the energy distributions of source events (colored crosses) and of the mean background in the three detectors black (solid histogram) before (left) and after (right) the background rejection. 
\begin{table}[!htb]
	\centering
	\begin{tabular}{c||cc}
		Parameter & Before rejection & After rejection \\   \hline\hline
		$n_{\rm H}$ & \multicolumn{2}{c}{$0.57^\dagger$ } \\ \hline
		$kT_{\rm BB}$ & 0.471$\pm 0.004$  &  0.472$\pm 0.004$ \\
		norm & 0.00111$\pm 0.00004$  & 0.00109$\pm 0.00004$  \\ \hline
		$\Gamma$ & 3.68$\pm 0.05$ & 3.70$\pm 0.05$  \\
		norm & 0.121$\pm 0.008$ & 0.123$\pm 0.008$ \\ \hline
		f & \multicolumn{2}{c}{1.0$^\dagger$} \\
		f$_{2}$ & 0.965$\pm 0.003$ & 0.965$\pm 0.003$ \\
		f$_{3}$ & 0.858$\pm 0.003$ & 0.858$\pm 0.003$ \\\hline
		$\chi^2$ & 520.0 with 441 d.o.f. & 529.5 with 441 d.o.f.\\ \hline
	\end{tabular}
	\caption{Results of \textsc{Xspec} fit of 4U 0142+61 data before and after the rejection, by using the spectral model: \textsc{TBabs*(BBody+PowerLaw)}, with normalization factors for the IXPE telescopes two and three. $^\dagger$Parameters value fixed in the spectral fit.}
	\label{tab:xspec_res}
\end{table}

The distributions are superimposed with the best-fit results obtained taking into account a background-subtraction in the IXPE nominal energy interval 2–8 keV: at higher energy, the background is dominant with respect to the source, but after rejection, this effect is strongly reduced. A comparison of the background level before and after the rejection procedure is shown in Figure \ref{fig:fit_xspec}--right. The fit results are reported in Table \ref{tab:xspec_res} for both original and background-rejected data sets.
The fit before and after background rejection show very well compatible results, moreover the reduced chi-square is also very similar $\chi^2/dof = 1.18$ before the rejection and 1.20 after it.

However, since the main scientific results for IXPE arise from the measurement of polarization, a joint spectro-polarimetric analysis of the observed data is performed by building a binned (40 eV) spectra for the Stokes parameters I, Q and U, which are simultaneously fitted with the usual \textsc{xspec} procedure of forward folding and applying a background subtraction. Polarization in a given energy interval was calculated by assuming the spectral model for I equal to TBABS*(BBODY+POWERLAW) with parameters fixed at the ones of Table \ref{tab:xspec_res} and convolving it with the constant polarization model \textsc{POLCONST} provided by \textsc{XSPEC}. Results from this procedure show a polarization PD = $(13.7 \pm 1.1)\%$ with PA = $(48 \pm 2)^\circ$ before the background rejection and PD = $(13.6 \pm 1.1)\%$ with PA = $(48 \pm 2)^\circ$ after it. 
\begin{figure}[!htb]
	\centering
	\includegraphics[width=0.75\textwidth]{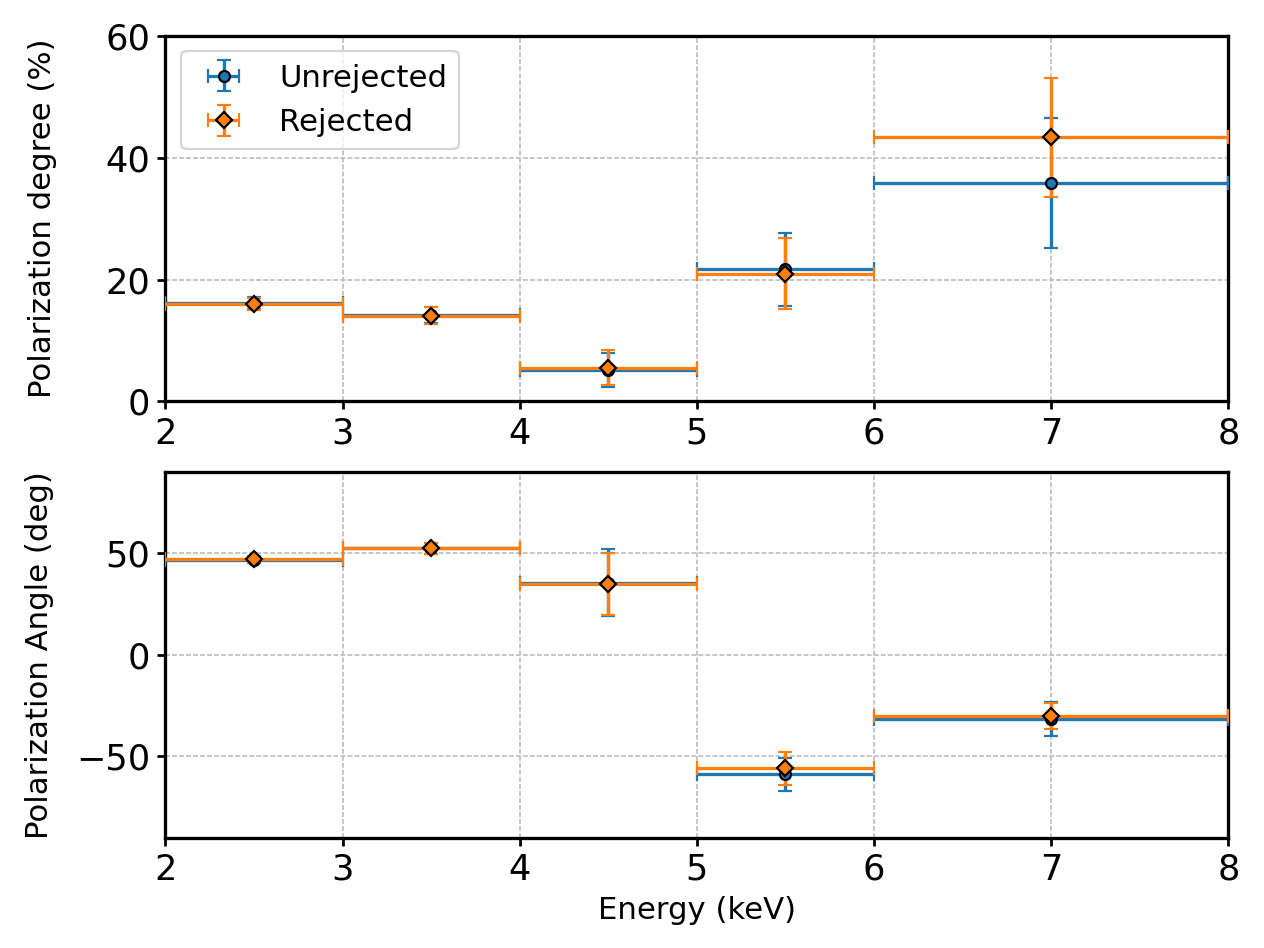}
	\caption{Polarization degree (top) and angle (bottom) obtained for 4U 0142+61 data fitted with \textsc{xspec} in different energy bins before (light blue) and after (orange) the background rejection. It is possible to observe that in the last energy bin, where the background dominates, the rejection strategy produces a more significant polarization degree because the dilution effect is partially washed out, while the polarization angle is well compatible in the two cases.}
	\label{fig:spe_pol}
\end{figure}
Repeating this analysis for different energy bins produces the result of Figure \ref{fig:spe_pol}, where it is possible to confirm that results are well compatible. A remarkable result arising from the background rejection is the improvement of significance for the polarization degree in the last energy bin (6--8 keV), where the background dominates. After the rejection, the background dilution effect is partially washed out, and the polarization degree increases from $\sim36$\% up to $\sim43$\%. Polarization degrees and angles are summarized in Table \ref{tab:pd_ener}.
\begin{table}[!htb]
    \centering
    \begin{tabular}{l|c|c|c|c|c|c||c}
    \multicolumn{2}{c|}{} & 2--3 keV & 3--4 keV & 4--5 keV & 5--6 keV & 6--8 keV & 2--8 keV \\\hline
    Unrejected & PD (\%) & $16.1\pm1.0$ & $14.2 \pm 1.4$ & $5\pm3$ & $22 \pm 6$ & $36\pm11$ & $13.7 \pm 1.1$ \\
    & PA ($^\circ$) & $46.9\pm1.8$ & $52 \pm 3$ & $36 \pm 17$ & $-59 \pm 8$ & $-32 \pm 8$ & $48\pm2$\\
    & $\sigma$ & 16 & 10 & 0.54 & 3.6 & 1.5 & 12 \\\hline
    Rejected &  PD (\%) & $16.1\pm1.0$ & $14.1 \pm 1.4$ & $6\pm3$ & $21 \pm 6$ & $43\pm10$  & $13.6 \pm 1.1$\\
    & PA ($^\circ$) & $47.0\pm1.8$ & $52 \pm 3$ & $35 \pm 15$ & $-56 \pm 8$ & $-30 \pm 6$ & $48\pm2$\\
     & $\sigma$ & 16 & 10 & 0.78 & 3.6 & 2.8 & 12 \\\hline
    \end{tabular}
    \caption{Polarization degree and angle obtained fitting by using \textsc{xspec} 4U 0142+61 data in different energy intervals within the IXPE nominal energy band. The significance ($\sigma$) is obtained in a non-polarized test hypothesis, and its estimate is based on the standard normal distribution; hence the corresponding P value means the probability that such a polarized signal is generated by a non-polarized source.}
    \label{tab:pd_ener}
\end{table}

\section{Conclusion}

The IXPE detectors allow to select background events thanks to imaging capabilities, with the best selection being an annular region with an inner radius of 2.5' and an outer radius of 5'. Larger selections or circular ones can introduce systematics in the polarization because of the detector's geometrical effects.

The IXPE background is at a level of 0.003 cps/arcmin$^2$. In this paper, a possible rejection approach is presented, allowing to remove up to 40\% of background events. Such a rejection approach is more effective for faint sources because, for the brighter ones, the background region is dominated by source events scattered by mirrors. Because of this, three possible background treatments can be performed:
\begin{itemize}
	\item Bright sources (rate $>2$ cps/armin$^2$): background is negligible, rejection can be applied, but it is neither effective nor useful, and subtraction from the same field should not be performed because the background region is dominated by the source itself;
	\item Faint sources (rate $<1$ cps/armin$^2$): background rejection is recommended and effective: after rejection, the residual background contribution should be subtracted in the analysis;
	\item Intermediate sources: background rejection is recommended and effective, after the rejection, the background region is still dominated by source events, which means that subtraction in the analysis should be avoided unless using a template background extracted from different sources.
\end{itemize}

As an example, the 4U 0142+61 data have been analyzed using prescriptions on the background obtained in this study. The result shows that, when the background impact is negligible, the rejection results are well in agreement with the ones arising from unrejected data, but when the background is comparable or higher than the source counting rate, the background rejection becomes crucial to improve the significance of the result.

\section{Acknowledgements}

The Imaging X-ray Polarimetry Explorer (IXPE) is a joint US and Italian mission. The US contribution is supported by the National Aeronautics and Space Administration (NASA) and led and managed by its Marshall Space Flight Center (MSFC), with industry partner Ball Aerospace (contract NNM15AA18C). The Italian contribution is supported by the Italian Space Agency (Agenzia Spaziale Italiana, ASI) through contract ASI-OHBI-2017-12-I.0, agreements ASI-INAF-2017-12-H0 and ASI-INFN-2017.13-H0, and its Space Science Data Center (SSDC) with agreements ASI-INAF-2022-14-HH.0 and ASI-INFN 2021-43-HH.0, and by the Istituto Nazionale di Astrofisica (INAF) and the Istituto Nazionale di Fisica Nucleare (INFN) in Italy.  This research used data products provided by the IXPE Team (MSFC, SSDC, INAF, and INFN) and distributed with additional software tools by the High-Energy Astrophysics Science Archive Research Center (HEASARC), at NASA Goddard Space Flight Center (GSFC).

\bibliography{references.bib}{}
\bibliographystyle{aasjournal}

\end{document}